\documentclass[aerospace,article,submit,pdftex,moreauthors]{Definitions/mdpi} 

\firstpage{1} 
\pubvolume{1}
\issuenum{1}
\articlenumber{0}
\pubyear{2023}
\copyrightyear{2023}
\datereceived{} 
\dateaccepted{} 
\datepublished{}
\hreflink{https://doi.org/} 

\usepackage[utf8]{inputenc}
\usepackage[T1]{fontenc}
\usepackage{wrapfig}
\usepackage{multirow}
\usepackage{mathtools}
\usepackage{booktabs}       
\usepackage{amsfonts}       
\usepackage{amsmath}
\usepackage{nicefrac}       
\usepackage{microtype}      

\usepackage{pdfpages}

\usepackage[normalem]{ulem} 

\DeclareMathOperator*{\argmax}{argmax}

\definecolor{Gray}{gray}{0.9}
\definecolor{Pink}{HTML}{FFA07A} 
\definecolor{Blue}{HTML}{728ed6} 

\Title{Lessons Learned in ATCO2: 5000 hours of Air Traffic Control Communications for Robust Automatic Speech Recognition and Understanding}

\TitleCitation{Lessons Learned in ATCO2: 5000 hours of Air Traffic Control Communications for Robust Automatic Speech Recognition and Understanding}

\Author{Juan Zuluaga-Gomez $^{\star, 1,2}$\orcidA{}, Iuliia Nigmatulina$^{1,3}$, Amrutha Prasad$^{1,4}$, Petr Motlicek$^{1,4}$\orcidB{}, Driss Khalil$^{1}$, Srikanth Madikeri$^{1}$, Allan Tart$^{5}$, Igor Szoke$^{4}$, Vincent Lenders$^{6}$, Mickael Rigault$^{7}$, Khalid Choukri$^{7}$
}

\AuthorNames{Juan Zuluaga-Gomez, Iuliia Nigmatulina, Amrutha Prasad, Petr Motlicek, Driss Khalil, Srikanth Madikeri, Allan Tart, Igor Szoke, Vincent Lenders, Mickael Rigault, Khalid Choukri}

\address{$^{1}$ \quad Idiap Research Institute, Martigny, Switzerland\\
$^{2}$ \quad Ecole Polytechnique F\'ed\'erale de Lausanne, Switzerland \\
$^{3}$ \quad University of Zurich, Switzerland\\
$^{4}$ \quad Brno University of Technology, Brno, Czechia\\
$^{5}$ \quad OpenSky Network, Burgdorf, Switzerland\\
$^{6}$ \quad Cyber-Defence Campus, armasuisse, 3602 Thun, Switzerland\\
$^{7}$ \quad Evaluations and Language Resources Distribution Agency (ELDA), Paris, France\\
}

\corres{Correspondence: juan-pablo.zuluaga@idiap.ch}

\abstract{
Voice communication between air traffic controllers (ATCos) and pilots is critical for ensuring safe and efficient air traffic control (ATC). This task requires high levels of awareness from ATCos and can be tedious and error-prone. Recent attempts have been made to integrate artificial intelligence (AI) into ATC in order to reduce the workload of ATCos. However, the development of data-driven AI systems for ATC demands large-scale annotated datasets, which are currently lacking in the field. This paper explores the lessons learned from the ATCO2 project, a project that aimed to develop a unique platform to collect and preprocess large amounts of ATC data from airspace in real time. Audio and surveillance data were collected from publicly accessible radio frequency channels with VHF receivers owned by a community of volunteers and later uploaded to Opensky Network servers, which can be considered an “unlimited source” of data. In addition, this paper reviews previous work from ATCO2 partners, including (i)~robust automatic speech recognition, (ii)~natural language processing, (iii)~English language identification of ATC communications, and (iv)~the integration of surveillance data such as ADS-B. We believe that the pipeline developed during the \textit{ATCO2 project}, along with the open-sourcing of its data, will encourage research in the ATC field. A sample of the ATCO2 corpus is available on the following website: \url{https://www.atco2.org/data}, while the full corpus can be purchased through ELDA at \url{http://catalog.elra.info/en-us/repository/browse/ELRA-S0484/}. We demonstrated that ATCO2 is an appropriate dataset to develop ASR engines when little or near to no ATC in-domain data is available. 
For instance, with the CNN-TDNNf kaldi model, we reached the performance of as low as 17.9\% and 24.9\% WER on public ATC datasets which is 6.6\% and 7.6\% better than with "out-of-domain" but supervised CNN-TDNNf model, respectively. We hope that this paper will contribute to the advancement of data-driven AI systems in ATC and ultimately lead to safer and more efficient AI-tools for ATC.
}

\keyword{Air traffic control communications; Automatic speech recognition and understanding; Opensky Network; Callsign recognition; ADS-B data.}

\begin{document}

\section{Introduction}

There has been a growing interest in the development of automatic speech recognition (ASR) and understanding systems for air traffic control (ATC) due to their potential to enhance the safety and efficiency of the aviation industry. The application of ASR and understanding technologies in ATC has resulted in the creation of advanced proof-of-concept engines that can assist air traffic controllers (ATCos) in their daily tasks. These engines are designed to analyze spoken ATC communications and convert them into machine-readable texts, allowing for faster and more accurate processing. Previous work such as MALORCA~\cite{kleinert2018semi,kleinert2019adaptation,helmke2016reducing,helmke2017increasing}, HAAWAII~\cite{nigmatulina2022two,kleinert2021automated}, or StarFish~\cite{kleinert2021apron} have shown mature-enough engines to reduce ATCos' workload\footnote{A clear example is described in~\cite{helmke2016reducing}. The authors concluded that integrating novel ASR-based tools can reduce the total amount of time that ATCos expend on entering and confirming the clearances in their workstations by 20\% absolute points.}
while increasing safeness. As a result, ASR and understanding technologies are becoming more advanced and capable of handling the complexities of ATC communications, leading to improved safety and efficiency in the aviation industry. The main caveat is that all of these engines are designed for specific solutions, i.e., one airport. The process of adapting AI-based engines to different airports or control areas needs new in-domain data and stays a challenge. For instance, ATC audio data collected from one airport, e.g., \texttt{airport $X$}, in general, does not transfer successfully to another airport, e.g., \texttt{airport $Y$}.

\begin{figure}[t]
    \centering
    \includegraphics[width=0.9\linewidth]{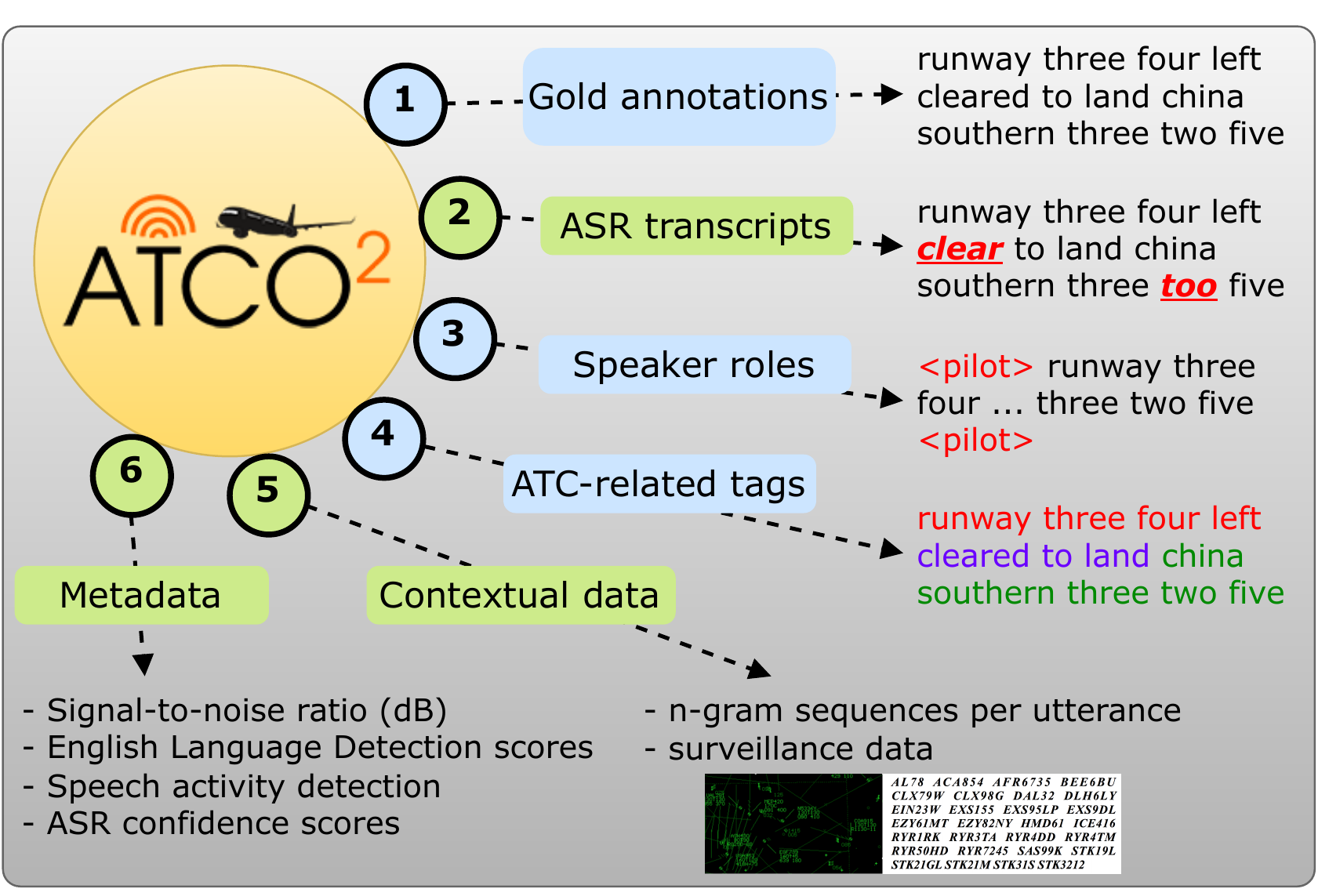}
    \caption{\textit{ATCO2 corpus ecosystem}. Blue circles denote annotations only available for \textit{ATCO2 test set corpus}. Green circles denote annotations and metadata available for both \textit{ATCO2 test set} and \textit{ATCO2 pseudo-labeled} corpus sets. Taken from previous work in~\cite{zuluaga2022atco2}.}
    \label{fig:atco2-ecosystem}
\end{figure}

One of the biggest bottlenecks faced by the community working on ASR for ATC is the collection and annotation of audio data. This process is highly expensive and demanding, requiring at least eight hours of man-hours to annotate one hour of ATC speech~\cite{cordero2012automated,ferreiros2012speech}. For some solutions, such as those targeting small airports, this cost may be prohibitive. Thus, it raises the question of what is the most efficient manner to collect and process more ATC audio data. Fortunately, recent projects like ATCO2 aim to overcome this obstacle~\cite{zuluaga2022atco2}. An overall overview of the ATCO2 corpora ecosystem is in Figure~\ref{fig:atco2-ecosystem}.

The ATCO2 project aims to reduce the human effort required to collect, pre-process, and transcribe ATC voice communications by employing state-of-the-art ASR and natural language processing (NLP) systems~\cite{kocour2021automatic,zuluaga2021bertraffic}. ATCO2 has released the largest corpus of ATC voice communications to date, consisting of more than 5,000 hours of automatically transcribed audio data and the correspondent to it surveillance data~\cite{kocour2021automatic}. In addition, four hours of human-transcribed data (i.e., gold annotations) have also been released. Through the ATCO2 project, we have discovered that the transcription process can be drastically sped up by providing data annotators with automatically transcribed data, rather than requiring them to generate annotations from scratch. This has reduced the real-time factor (RTF) of the annotating process by 30 points, from 50 to 20 RTF~\cite{zuluaga2022atco2}.

This paper covers several aspects and lessons learned related to the data collection and annotation pipeline, primary actors for collection and annotation, and the main AI-based systems developed during ATCO2. It also covers baselines on ATC-related ASR and understanding and novel techniques, e.g., integration of surveillance data into the ASR pipeline~\cite{kocour21_interspeech,nigmatulina2021improving,nigmatulina2022two}.

The rest of the paper is organized as follows.
Section~\ref{sec:related-work} covers some related work on automatic speech recognition and understanding for ASR. 
We describe the ATCO2 ecosystem, data collection pipeline and some other characteristics in Section~\ref{sec:atco2-corpus}. In Section~\ref{sec:platform-volunteers} we cover technical details about the data collection platform (front end and back end) and how the community of volunteers interacts with it. In Section~\ref{sec:technologies-atco2} we cover the main technologies that can be developed with ATCO2 corpora. Finally, we provide brief future work directions 
 and conclude the paper in Section~\ref{sec:conclusion}.

\section{Early Work on ATC}
\label{sec:related-work}

Recent work on ASR and understanding of ATC communications has been documented in~\cite{tarakan2008automated,ferreiros2012speech}. In recent years, more research-oriented work has focused on pure ASR. For example,~\cite{zuluagagomez20_interspeech} established the first benchmark on ASR for different ATC communications-focused databases. Furthermore, there has been a significant effort to integrate novel semi-supervised learning algorithms for boosting the ASR performance with surveillance data such as~\cite{srinivasamurthy2017semi,kleinert2018semi,zuluagagomez21_interspeech}. This supports the idea of the growing interest in research in ASR and understanding towards ATC, with mature proof-of-concept engines that can assist ATCos in their daily tasks. Our previous work related to the large-scale automatic collection of ATC audio data from different airports worldwide is covered in~\cite{kocour2021automatic,zuluaga2020automatic}. Additionally, recent work targeted to improve callsign recognition by integrating surveillance data into the pipeline has been explored in~\cite{kocour21_interspeech,nigmatulina2021improving,nigmatulina2022two}.

Another line of work has been directed at open-sourcing ATC-related databases: for US-based communications~\cite{LDC_ATCC}, in Czechia~\cite{UWB_ATCC}, and~\cite{HIWIRE} for several accents in English. Recently, there was an Airbus-led challenge for ATC communications, with French-accented recordings from France~\cite{AIRBUS,pellegrini2018airbus}. Private databases were also explored in~\cite{vocalise,enac}. For a general overview of ATC-related databases, we redirect the reader to Table 1 in the previous work in~\cite{zuluaga2022atco2} and for the databases released by the ATCO2 project to Table~\ref{tab:databases}.

\begin{table}[t]
    \caption{Air traffic control communications corpora released by ATCO2 project. $^\dagger$full database after silence removal. $^{\dagger\dagger}$speaker accents depend on the airport's location; accents of pilots are not known at any time of the communication due to privacy regulations.}
    \label{tab:databases}
    \centering

    \resizebox{0.98\textwidth}{!}{
    \begin{tabular}{l p{6cm} c p{2.5cm} ll} \toprule
        \multicolumn{1}{l}{\textbf{Database}} & \textbf{Details} &
        \textbf{Licensed} &
        \textbf{Accents} & \textbf{Hours}$^\dagger$ & \textbf{Ref} \\ 
        \midrule
        \rowcolor{Gray}
        \multicolumn{6}{c}{\textbf{\textit{Released corpora by ATCO2 project}}} \\
        \midrule
        \textbf{\textit{ATCO2 corpora}} & Data from different airports and countries: public corpora \url{catalog.elra.info/en-us/repository/browse/ELRA-S0484/} & & Several$^{\dagger\dagger}$ & & \cite{zuluaga2022atco2} \\
        
        \cline{2-6}\rule{0pt}{3ex}$\hookrightarrow$ \textit{ATCO2-test-set} & Real life data for ASR and NLP research. & \checkmark & $\hookrightarrow$ & 4 & \cite{zuluaga2022atco2} \\

        $\hookrightarrow$ \textit{ATCO2-PL set} & Pseudo-labeled real data for research in ASR and \mbox{NLU}.  & \checkmark & $\hookrightarrow$ & 5281 & \cite{kocour2021automatic,zuluaga2022atco2} \\

        \cline{2-6}
        \rule{0pt}{3ex} & \multicolumn{5}{c}{\cellcolor{Gray}\textbf{\textit{Free access databases released by ATCO2 project}}} \\
        \cline{2-6}
        \rule{0pt}{3ex}$\hookrightarrow$ \textit{ATCO2-test-set-1h} & 'ASR dataset': public 1 hour sample, a subset of \textit{ATCO2-test-set}. \url{https://www.atco2.org/data} & \checkmark & $\hookrightarrow$ & 1 & \cite{kocour2021automatic} \\

        $\hookrightarrow$ \textit{ATCO2-ELD set} & 'ELD dataset': public dataset for English language detection. \url{https://www.atco2.org/data} & \checkmark & $\hookrightarrow$ &  26.5 & \cite{szoke21_interspeech} \\
        
        \bottomrule
    \end{tabular}
    }
\end{table}

\section{ATCO2 corpora}
\label{sec:atco2-corpus}

It is well known that AI-based tools need large amounts of reliably transcribed data during their training process. For instance, ASR or NLP tools for ATC could work better if we had large-scale data. The ATCO2 corpus was designed to target this data scarcity issue by solving three big challenges: \\

\textbf{(1) Current corpora related to air traffic control are primarily focused on automatic speech recognition}. However, for an AI engine to be successfully deployed in the control room, it must not only accurately transcribe ATC communication but also understand it. This includes the ability to detect speaker roles (SRD) as well as extract and parse callsigns and commands. The ATCO2 corpus provides a comprehensive solution to this challenge by including detailed tags for SRD and callsign and command extraction. This, in turn, will improve the accuracy and efficiency of AI-based systems in ATC operations.

\textbf{(2) Out-of-domain ASR and NLP-based corpora transfer poorly to the ATC domain.} ATC communication follows a unique grammatical structure and employs a specific set of the vocabulary defined by ICAO~\cite{allclear}, making it a niche application. This poses a significant limitation to the use of out-of-domain corpora.\footnote{Previous studies~\cite{zuluagagomez20_interspeech} have shown that employing non-ATC related corpora such as LibriSpeech~\cite{panayotov_librispeech_ICASSP2015} or others~\cite{ardila2019common,godfrey1992switchboard} does not match the acoustics of ATC communication, and therefore does not contribute to ASR training.} As such, the ATCO2 project collected and publicly released a large amount of ATC-specific data to aid in the development of ASR and understanding engines for ATC.

\textbf{(3) The research community working on ATC is hindered by a severe lack of openly available annotated data.} To address this issue, the ATCO2 project has released a vast corpus of over 5000 hours of automatically annotated data (i.e., \textit{ATCO2-PL set}), as well as 4 hours of manually annotated data (i.e., \textit{ATCO2-test-set-4h}), as shown in Table~\ref{tab:databases}.\footnote{It is worth noting that the annotations generated by the automatic tools have been proven to be robust, with word error rates (WER) as low as 9\% achieved when training an ASR engine using these transcripts alone, as shown in the \texttt{Malorca-Vienna-test} set from the MALORCA project~\cite{zuluaga2022atco2}.}

\begin{figure}[t]
    \centering
    \includegraphics[width=0.95\textwidth]{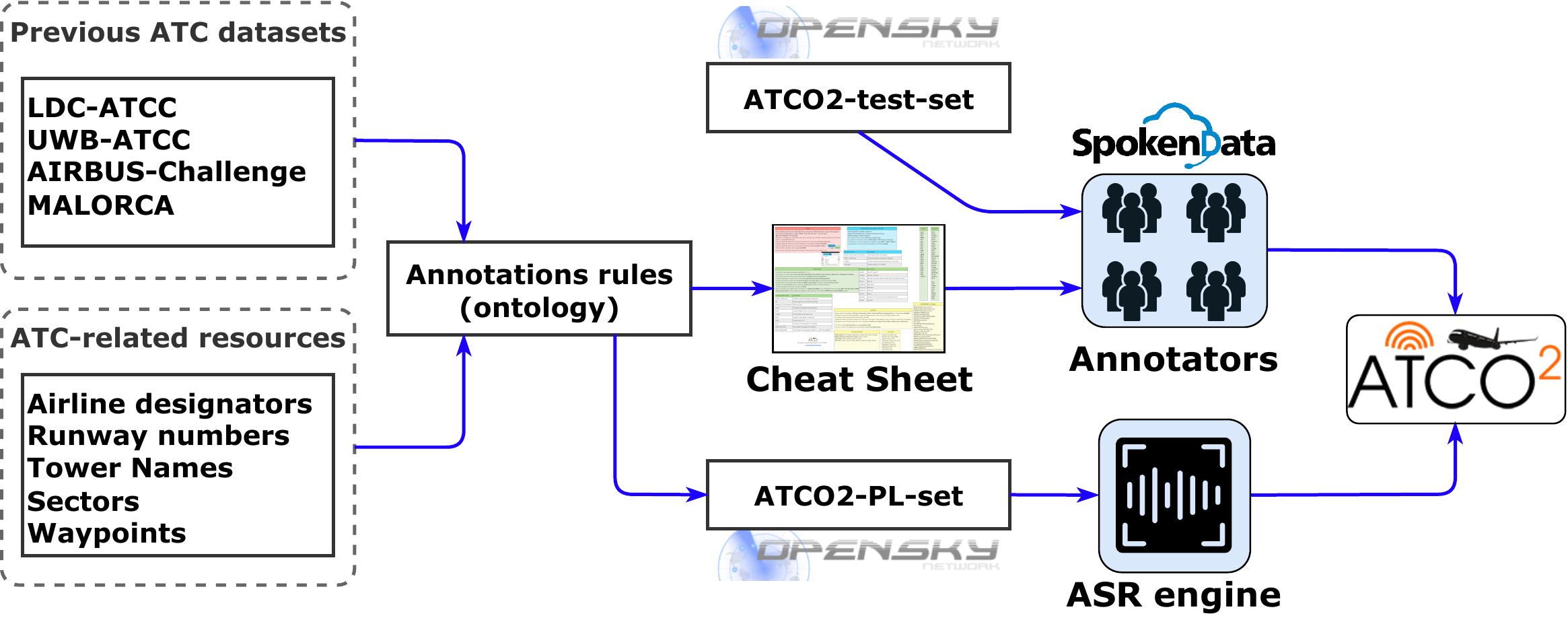}
    \caption{\textit{Annotation protocol workflow}. ATCO2 corpora follow a rigorous annotation protocol based on previous ATC-related corpora and resources. Additionally, a cheat sheet for the annotation of ATC transcripts was developed during the project. The cheat-sheet or annotation is available in Appendix~\ref{appendix:cheat-sheet}.}
    \label{fig:annotation-protocl}
\end{figure}

\subsection{ATCO2 Ecosystem and Generalities}

The ATCO2 ecosystem is described in Figure~\ref{fig:atco2-ecosystem}.
During the collection of the ATCO2 corpora, we followed several pre-processing steps in order to normalize the generated annotations. Here, we aim at minimizing errors produced by phonetic dissimilarities, e.g., \textbf{\textcolor{blue}{“descent to
two thousand"}} and \textbf{\textcolor{red}{“descend two two thousand"}}. We performed several text normalization steps in order to unify the gold and automatic annotations following ICAO rules~\cite{allclear} and well-known ontologies for ATC communications~\cite{helmke2018ontology}. A summary of the annotation protocol is depicted in Figure~\ref{fig:annotation-protocl}. Additionally, we direct the reader to a more detailed overview on text normalization and lexicon for data annotation in Section 3 of~\cite{zuluaga2022atco2}. The ATCO2 corpora are composed of \textit{ATCO2-PL-set} corpus and \textit{ATCO2-test-set} corpus, described below:

\begin{itemize}
    \item First, the \textbf{\textit{ATCO2-PL-set corpus}} is the first ever release of a large-scale dataset targeted to ATC communications. We recorded, pre-processed, and automatically transcribed $\sim$5281 hours of ATC speech from ten different airports, see Table~\ref{tab:stats-databases}. To the best of the author's knowledge, this is the largest and richest dataset in the area of ATC ever created that is accessible for research and commercial use. Further information and details are in~\cite{zuluaga2022atco2}.
    \item Second, \textit{\textbf{ATCO2-test-set-4h corpus}} was built for the evaluation and development of automatic speech recognition and understanding systems for English ATC communications. This dataset was annotated by humans. There are two partitions of the dataset as stated in Table~\ref{tab:databases}. The \textit{ATCO2-test-set-1h corpus} is an $\sim$1-hour long open-sourced corpus, and it can be accessed for free in: \url{https://www.atco2.org/data}. The \textit{ATCO2-test-set-4h corpus} contains \textit{ATCO2-test-set-1h corpus} and adds to it $\sim$3 more hours of manually annotated data. The full corpus is available for purchase through ELDA in: \url{http://catalog.elra.info/en-us/repository/browse/ELRA-S0484}. 
\end{itemize}

\begin{table}[t]
    \caption{Total accumulated duration (in hours) of speech after voice activity detection per airport in \textit{ATCO2-PL-set}. $^\dagger$English Language Detection (ELD) [0-1] score. This score tells how confident is our ELD system in detecting whether there is only English spoken inside the ATC communication.}
    \label{tab:stats-databases}
    \centering
    \begin{tabular}{ccccccccccc}
        \toprule
        \rowcolor{Gray}        
        \multicolumn{11}{c}{\textbf{\textit{ICAO (Airport) - City}}} \\ 
        \midrule
        \rotatebox{90}{EETN - Tallinn} & \rotatebox{90}{EPLB - Lublin} & \rotatebox{90}{LKPR - Prague} & \rotatebox{90}{LKTB - Brno} & \rotatebox{90}{LSGS - Sion} & \rotatebox{90}{LSZB - Bern} & \rotatebox{90}{LSZH - Zurich} & \rotatebox{90}{LZIB - Bratislava} & \rotatebox{90}{YBBN - Brisbane} & \rotatebox{90}{YSSY - Sydney} & \rotatebox{90}{others - others} \\
        \midrule
        \rowcolor{Gray}        
        \multicolumn{11}{c}{\textbf{\textit{English Data (language score $\geq 0.5$)$^\dagger$}}} \\  
        \midrule
        131 & <1 & 1762 & 888 & 330 & 699 & 921 & 24 & 170 & 77 & <1 \\
        \midrule
        \rowcolor{Gray}        
        \multicolumn{11}{c}{\textbf{\textit{Non-English Data (language score $< 0.5$)$^\dagger$}}} \\  
        \midrule        
        2 & <1 & 187 & 611 &  83  &  55  & 49 & 26  & 10 & 3 & <1 \\
        \bottomrule    
    \end{tabular}
\end{table}

\subsection{Data Collection Pipeline}
\label{subsec:data-collection-pipeline}

\begin{figure}[t!]
    \centering
    \includegraphics[width=0.95\textwidth]{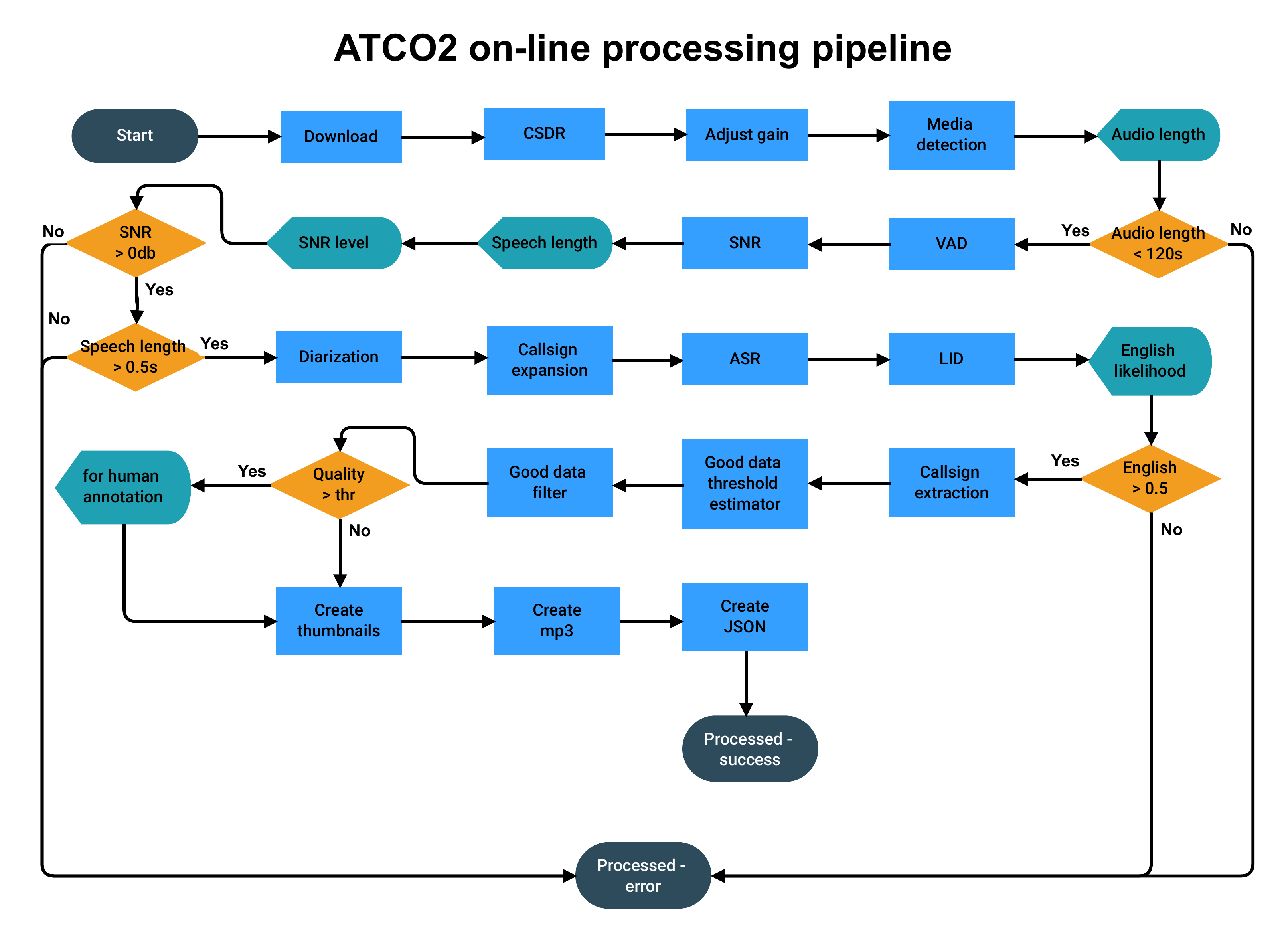}
    \caption{\textit{ATCO2 workflow for processing data collected by a community of feeders}. Initially, the data is sent and stored on OSN servers. The audio data goes through several modules to filter-out recordings with a high-level of noise and too long or too short segments. 
    \textbf{\textcolor{blue}{Blue rectangles}} are processes. The \textbf{\textcolor{cyan}{cyan arrow blocks}} are internal callback events, where the pipeline informs the master node about progress and sends intermediate results. The \textbf{\textcolor{orange}{orange rhombuses}} are conditions, where intermediate results are taken into account (e.g., an SNR level), i.e., whether to continue (clean audio) or stop processing. A final internal callback is run when the pipeline finishes. It triggers the API to call the OSN server with the particular callback, for instance, the processing has finalized as \textit{OK} or \textit{ERROR}.
    }
    \label{fig:pipeline}
\end{figure}

The processing pipeline is implemented as a Python script that follows a configuration file $\rightarrow$ \texttt{worker.py}. The configuration file allows us to modify the logic and flow of the data in the pipeline on-the-fly. It allows parallelism, forking, and conditions. In principle, \texttt{worker.py} consists of global definitions (constants), blocks (local definitions), and links (an acyclic oriented graph) between blocks. The processing pipeline is given in Figure~\ref{fig:pipeline}. For instance, we address earlier implementations of each technology from the previous work~\cite{kocour2021automatic,zuluaga2020automatic}, e.g., segmentation and diarization, ASR, or named entity recognition (NER). All the technologies and tools are encapsulated in a \texttt{BASH scripts} with a unified interface. 

The first row of blocks from Figure~\ref{fig:pipeline} refers to segmentation and demodulation. Initially, an antenna and a recording device jointly capture the radio signal, which is divided into segments containing portions where the transmission was “active”, the silent parts are not recorded (push-to-talk is used in ATC voice communication). This functionality is part of the RTLSDR-Airband audio recording software, from which we dump the raw I/Q signal. Second, we convert this complex I/Q radio signal into a waveform signal by a software defined radio CSDR. The first part is done in the recording device, while the second is done at OpenSky Network (OSN).\footnote{The OpenSky Network is a non-profit community-based receiver network which has been continuously collecting air traffic surveillance data since 2013. Unlike other networks, OpenSky keeps the complete unfiltered raw data and makes it accessible to academic and institutional researchers.}

Next, we do “signal-to-noise ratio (SNR) filtering” (second row), the purpose is to remove the recordings that are too noisy. In bad recording conditions, we can end up in a situation in which the voice is not intelligible. The following step is “diarization” (third row). In the automatically segmented data, some recordings contain more than one speaker. This is a problem because we intend to automatically transcribe speaker turns of single speakers. And, for subsequent NLP/SLU tasks, it is important to separate the speaker turns as well. The diarization solves this by splitting the audio into segments with single-speakers and assigning them speaker labels. In the ASR step, we simply convert “speech-to-text”. This is done by our ASR system that we build with tools from the Kaldi toolkit~\cite{povey2011kaldi}. The output from this step are transcripts, which inevitably contain some errors. To improve the accuracy of the transcripts, we use contextual information (callsign lists from surveillance data\footnote{The callsign lists come from the air traffic monitoring databases of OpenSky Network.}). Further details in Section~\ref{subsec:boosting} and~\cite{kocour21_interspeech,nigmatulina2021improving}.

Next, the transcripts are used as input for the English language detection (ELD) system. The purpose is to be able to discard non-English audio data. The typical state-of-the-art language identification system is based on acoustic modeling and uses audio as input. For the ATC speech, we do not need to “identify” the non-English languages, so we developed a “lexical English detection system” which uses transcripts and confidence scores produced by ASR as its inputs.\footnote{This work is covered in our previous paper presented at Interspeech in 2021~\cite{szoke21_interspeech}.} For ATC speech, this worked better than the “traditional” acoustic language identification method. 
The last automatic operation is “post-processing by NLP”. Currently, the pipeline performs a Callsign-code Extraction step. It returns the callsign in ICAO format like “DLH77RM” belonging to an aircraft. 
Finally, some processed data goes through “human correction”, and some data is kept with automatic labels. The former case produced \textit{ATCO2-test-set-4h corpus}, while the latter, \textit{ATCO2-PL-set corpus}. A more detailed description of the data collection flow and data annotation is in Appendix~\ref{appendix:anno-data-flow} and in Appendix~\ref{appendix:communication-schema}.

\subsection{Quality Estimation for Data Annotation}

As mentioned at the end of Section~\ref{subsec:data-collection-pipeline}, the captured, processed, and automatically transcribed data (see Figure~\ref{fig:pipeline}) can be annotated by humans. This in turn would generate 'gold annotations' that we use to evaluate the proposed ASR and NLP systems. The annotation of \textit{ATCO2-test-set-4h corpus} went through all these steps. However, before annotation, we need to select the most intelligible and higher quality data for human transcriptions. As the data is continuously being recorded by OSN, we need to select the most intelligible and clean data. We developed a score that ranks the recordings depending on their quality. this score integrates seven metrics that assess the quality of each recording present in \textit{ATCO2 corpora}. For instance, we use Equation~\ref{eq:quality-audio} to measure, rank and select the ATC communications with the highest quality. Later, these recordings were shortlisted for human annotations (see Section~\ref{subsec:transcriber}). The data annotators generated the ground truth transcripts and tags that are part of the \textit{ATCO2-test-set-4h corpus}. The ranking score is given by:

\begin{equation}
    \label{eq:quality-audio}
    \begin{split}
        Score = \log(avg_{SNR} + e) + \log(num_{spk} + e) + \log\left(\dfrac{speech_{len}}{audio_{len}} + e\right) + \\
        ELD_{score} \times 3~+ avg_{WordConf} \times 3 + \log(wrd_{cnt} + e),
    \end{split}
\end{equation}

where,

\begin{itemize}
    \item $avg_{SNR}$ — provides average SNR of speech in range <0, 40>. We want SNR to be as high as possible,
    \item $num_{spk}$ — provides the number of speakers in the audio in the range of <1, 10>. The more speakers detected in audio, the better,
    \item $speech_{len}$ — provides the amount of speech in seconds,
    \item $audio_{len}$ — provides the overall audio length. More speech detected in audio is better,
    \item $ELD_{score}$ provides “probability” of audio being English in the range <0.0, 1.0>, the higher, the better,
    \item $avg_{WordConf}$ - provides average confidence of the speech recognizer <0.0, 1.0>. We want data where the recognizer is confident. Higher is better,
    \item $wrd_{cnt}$ - provides the number of words spoken in the range of <0, ~150>. The more words, the better.
\end{itemize}

\begin{figure}[t]
    \centering
    \includegraphics[width=0.95\linewidth]{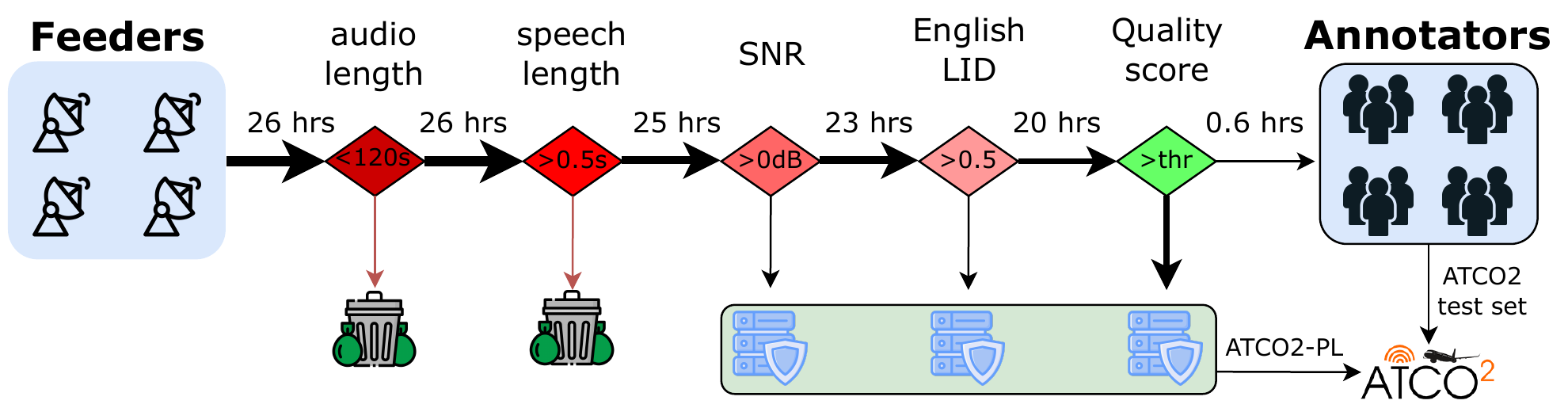}
    \caption{\textit{Breakdown of data flow yield from raw data (recordings from data feeders) to the human-annotated transcripts throughout the pipeline}. This is a one-day snapshot from 9.2.2022 (DD/MM/YYYY).}
    \label{fig:data_flow}
\end{figure}

A breakdown of the outputs of these steps for a single day is in Figure~\ref{fig:data_flow}. For instance, $\sim$0.6\,hrs of data are selected for gold annotations from an initial 26\,hr-pool of audio data. We believe this is a robust quality scoring method because it gathers information from different systems, e.g., ASR, SNR, and ELD estimation. A day-to-day estimation of the output of each of these steps is available on the SpokenData website: \url{https://www.spokendata.com/atc}.

\subsection{Runtime Characteristics}

We also measured the running time for individual components of our processing pipeline. In Table~\ref{tab:runtime-chars}, we list the relative time spent by each module, such as ASR and diarization. Other important parts are pre-processing, voice activity detection (VAD) segmentation, and ELD. The pre-processing involves: getting data, demodulation by software radio, segmental gain control, detecting media format, and plotting waveform. A key metric is the real-time factor of the whole pipeline. The real-time factor is the ratio of `processing time` over `length of the audio`. Our processing pipeline has a real-time-factor of 4.47. In other words, the processing is computationally demanding. For an average five-second-long recording, the processing time is 22 seconds. The actual running times of each component for the “average” five-second long recording are shown in Table~\ref{tab:runtime-chars}. 

\begin{table}[t]
    \caption{Division of processing time into components of the processing pipeline. The values in the second column are for an average 5.016 second long recording. The average was computed over 10334 recordings (14.4 hours), recorded on 04.12.2012 (DD/MM/YYYY).}
    \centering
    \label{tab:runtime-chars}
    \begin{tabular}{lll}
        \toprule
        \rowcolor{Gray} \textbf{Processing step} & \textbf{Time [s]} & \textbf{Percentage [\%]} \\
        \midrule
        Pre-processing & 2,540 & 11,62 \\
        VAD segmentation & 2,436 & 11,15 \\
        SNR estimation & 0,655 & 3,00 \\
        Diarization & 7,128 & 32,62 \\
        Callsign expansion & 0,464 & 2,12 \\
        Speech-to-text (ASR) & 7,021 & 32,13 \\
        English Detection & 1,292 & 5,91 \\
        Callsign extraction & 0,069 & 0,32 \\
        Post-processing & 0,245 & 1,12 \\
        \midrule
        \rowcolor{Gray} \textbf{Total time} & \textbf{21,850} & \textbf{100,00} \\
        \bottomrule
    \end{tabular}
\end{table}

\section{Collection Platform and Community of Volunteers}
\label{sec:platform-volunteers}

\begin{figure}[t]
    \centering
    \includegraphics[width=0.7\linewidth]{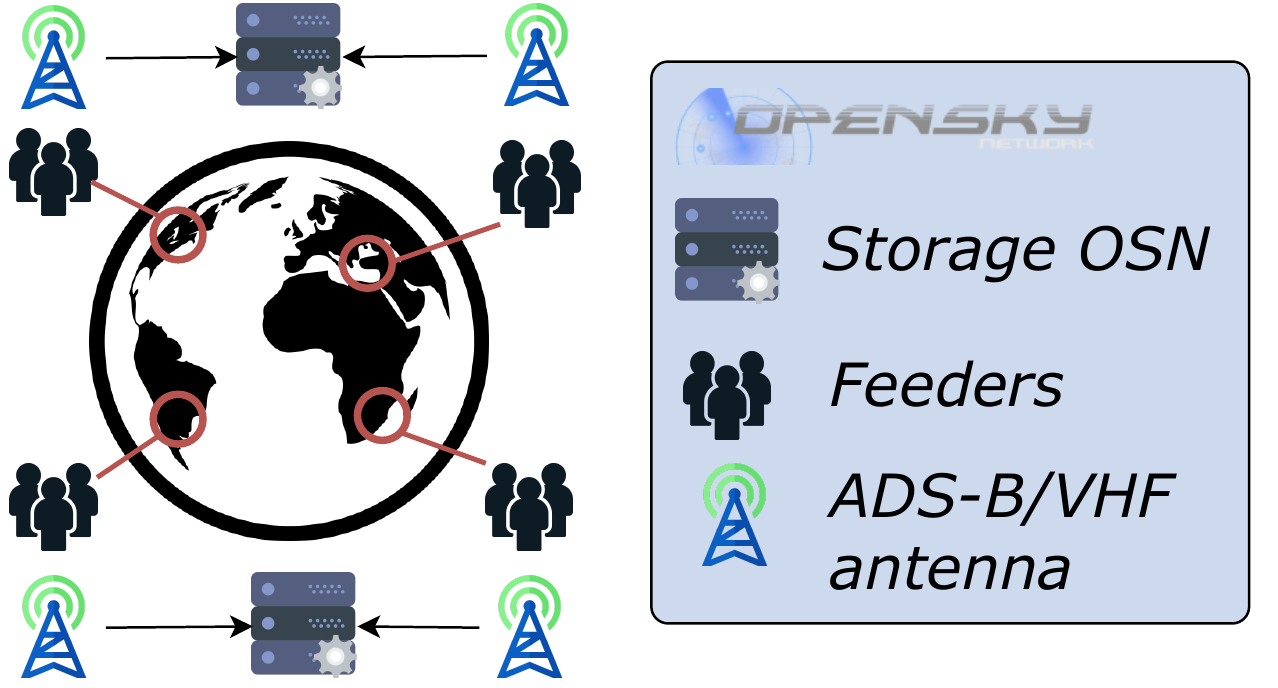}
    \caption{\textit{Data feeders pipeline}. The data users have set up a VHF receiver and feed data to OSN servers.}
    \label{fig:data-feeders}
\end{figure}

In this section, we summarize the data collection and distribution. In addition, a short description of the roles involved in data processing is provided. We also cover some high-level statistics about the collected data. 
First, data is captured and fed into the OpenSky Network by the volunteers who operate their own receiver equipment (see Figure~\ref{fig:data-feeders}). These individuals are often aviation enthusiasts with previous operational experience, or people with an interest in aviation technology, e.g., conducting domain-related research. But anyone with little-to-no background in aviation or technology can become a feeder. To become a feeder, one must have an internet connection and access to a VHF receiver. An affordable low-complexity setup is covered in the ATCO2 corpus paper~\cite{zuluaga2022atco2} and the guide for setting it up is provided \url{https://ui.atc.opensky-network.org/set-up}. It is important to recall that in some countries, it is prohibited by law to record air traffic management (ATM) data. For instance, we recommend checking the \textbf{Legal and privacy aspects for collection of ATC recordings} section in~\cite{zuluaga2022atco2}. Also, we redirect the reader to recent work related to legal aspects of ATC data collection in~\cite{rigault2022legal}.

\subsection{The Platform}

\begin{figure}[t]
    \centering
    \includegraphics[width=0.9\linewidth]{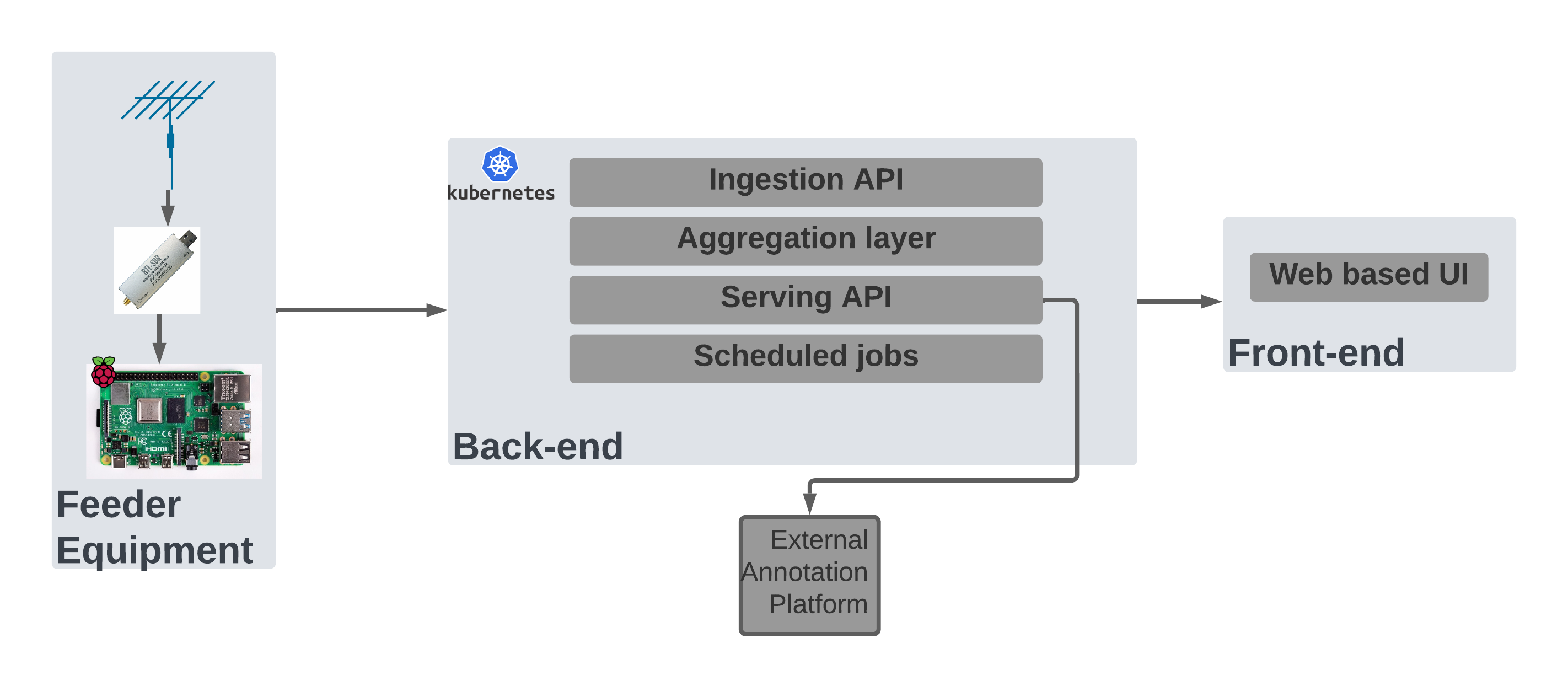}
    \caption{\textit{The high-level architecture of the data collection platform}.}
    \label{fig:osnDataCollectionPlatform}
\end{figure}

The high-level architecture is given in Figure~\ref{fig:osnDataCollectionPlatform}. As one can observe, the platform has been divided into three distinct groups: (i)~Feeder Equipment, (ii)~Back-end, and (iii)~Front-end. The architecture was decided during the design phase of ATCO2, with the main objective to achieve scalability of the entire system. That meant keeping the complexity relatively low within all the groups. The low complexity would allow to:

\begin{itemize}
    \item support a similar number of users as the current OpenSky ADS-B system; 
    \item keep the feeder equipment simple and affordable; 
    \item provide data to different types of users in a simple and intuitive way; 
    \item interface external services (e.g., voice annotation) in a simple and intuitive way; 
    \item keep maintenance and error handling as simple as possible.     
\end{itemize}

A better overview of the OSN platform is also listed in Appendix~\ref{appendix:transcription-pipeline}. As mentioned above, the platform has been divided into three parts. Below we describe each of these platform's groups. \\

\textbf{Feeder Equipment:} The main task of the feeder equipment is to capture the conversation between the pilot and the ATCo and feed the data, together with some relevant metadata to the back-end. For the recording part, we recommend using RTLSDR-Airband together with RTL-SDR dongle. It is an affordable and widely used combination within the aviation enthusiast community for this exact purpose—to capture and stream ATC voice. 

The feeder software is responsible for transmitting the recordings from the receiver to the remote server. It is a rather simple piece of software that monitors the output directory of the RTLSDR-Airband and transfers any new data it finds to the back-end using a GRPC connection. The fact that the feeder software only looks for specific types of data from the output folder suggests that the feeder is free to choose any other software for capturing and storing the voice data. Care must be taken to assure that the output is suitable for the feeder software. A simple, step-by-step guide has been provided to simplify the setup process. It can be found on \url{https://ui.atc.opensky-network.org/set-up}. \\

\textbf{Back-end:} The main tasks for the back-end are: (i)~to store recordings, transcripts, and any other relevant metadata and (ii)~to provide interfaces for external users. The external users in this are data feeders, transcription service providers, data users, or any other parties contributing to the dataset or making use of it. The back-end is deployed on Kubernetes, an open-source container orchestration system. As one can observe from Figure~\ref{fig:osnDataCollectionPlatform}, there are several processing layers involved. These layers are: 

\begin{itemize}
    \item Ingestion API: receives recording segments and metadata and queues them for processing in Kafka/S3 compatible object storage.
    \item Aggregation layer: converts raw data to flac audio, stores metadata, and triggers transcription using Kafka Streams, S3, and Serving API.
    \item Serving API: provides external interfaces to consume metadata, store, and consume transcript and stats.
    \item Scheduled jobs: run processes that are not part of the streaming process like stats aggregation and data housekeeping. 
\end{itemize}

Interfacing the back-end is done using API, which is well documented in \url{https://api.atc.opensky-network.org/q/swagger-ui}. In order to access the Back-end and make use of the available API-s, one needs to register on \url{https://auth.opensky-network.org/auth/}, contact OpenSky Network (mailto: \texttt{contact@opensky-network.org}), and give a short description of what one needs the access for. \\

\textbf{Front-end:} Front-end is a web-page (\url{https://ui.atc.opensky-network.org/}) and it provides access to public stats, links to documentation (e.g., API documentation), and external web-pages (e.g., SpokenData transcription service). In addition, this is a place for a user to set up their receivers, see some statistics about the receiver performance, and so on. 

\subsubsection{Statistics}

\begin{figure}[t]
    \centering
    \includegraphics[width=0.7\linewidth]{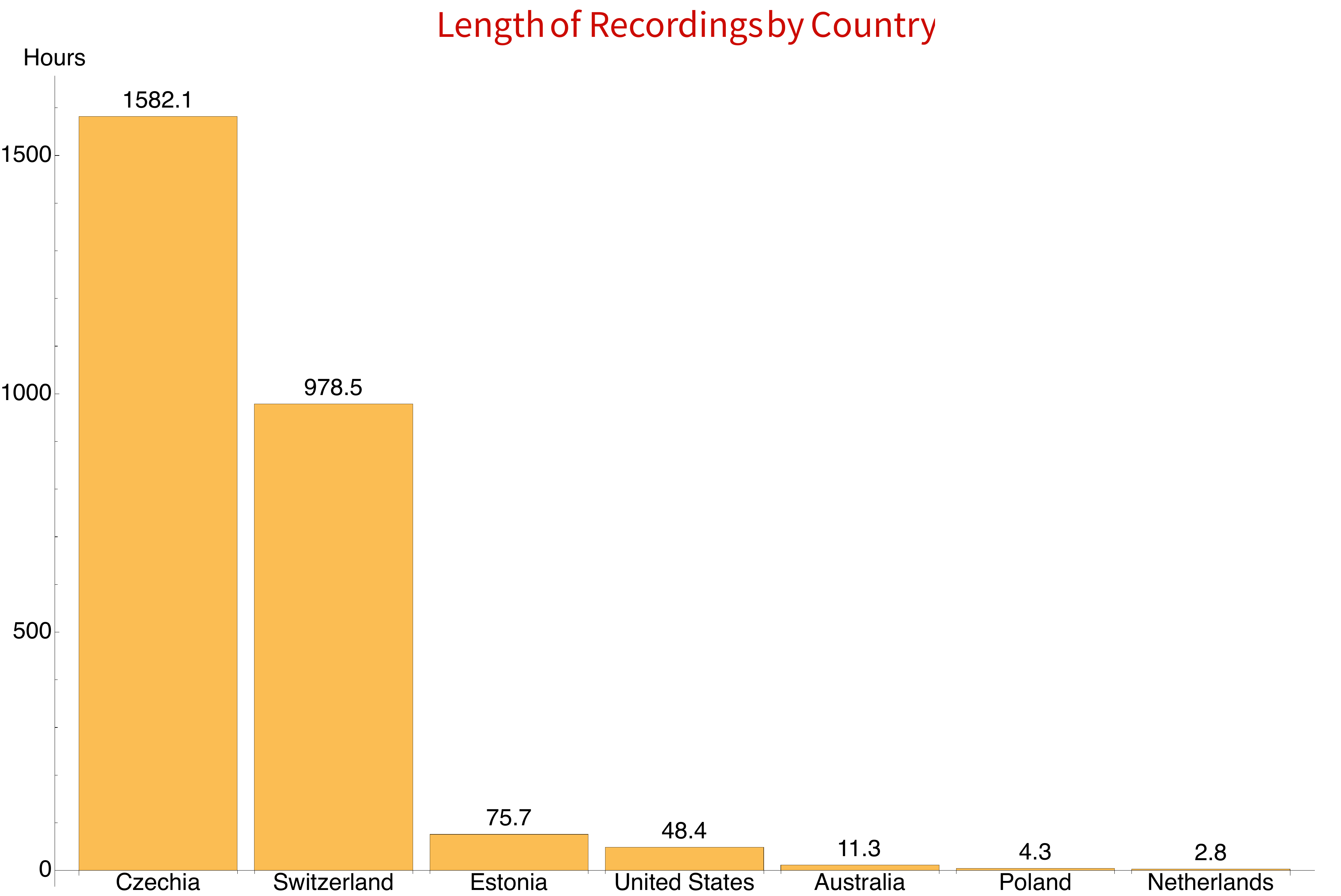}
    \caption{\textit{Length of recordings per country from the beginning of the service until March 5th, 2023}. Countries, where the length of recordings is longer than 1 hour given. }
    \label{fig:recrodingLengthCountry}
\end{figure}

\begin{figure}[t]
    \centering
    \includegraphics[width=0.7\linewidth]{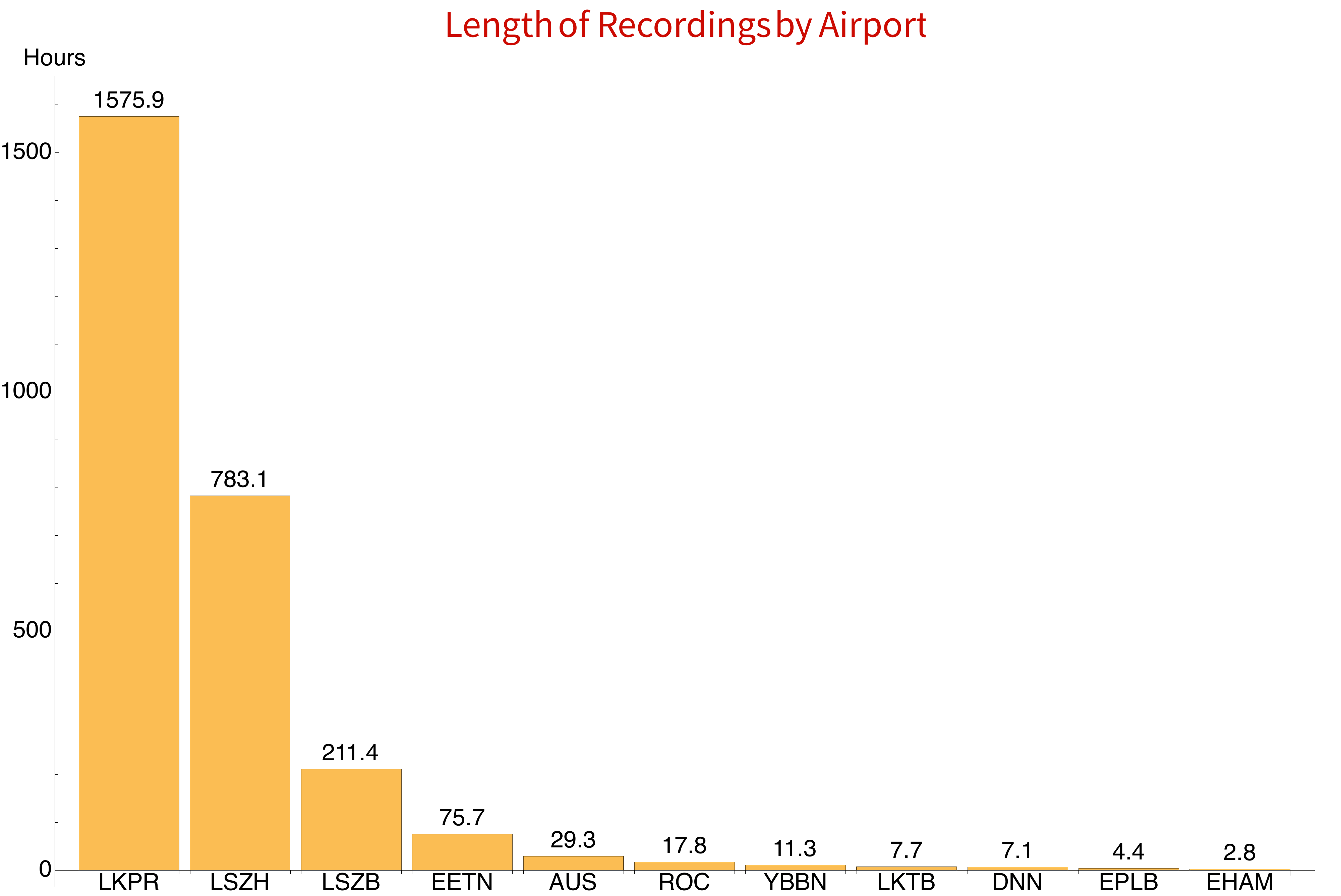}
    \caption{\textit{Length of recordings per airport from the beginning of the service until March 5th, 2023}. Airports, where the length of recordings is longer than 1 hour given. }
    \label{fig:recrodingLengthAirport}
\end{figure}

Since the opening of the service (March 5th, 2023), the ATCO2 project has recorded speech from 24 different airports in 14 different countries. In Figure~\ref{fig:recrodingLengthCountry} and Figure~\ref{fig:recrodingLengthAirport}, names of countries and airports together with corresponding recording lengths are shown. Please note that only the places with the length of regrinding $\geq 1$ hours are included. 

\subsection{Data Annotators}
\label{subsec:transcriber}

Apart from the data feeder, there is another type of volunteers who have contributed to the project and will continue to contribute in the future. These are called 'Annotators'. The data annotators are volunteers who write down the transcripts of the ATC voice communications, including assigning speakers and labeling named entities (such as callsigns and commands). For the ATCO2 project, we relied on both volunteers and paid transcribers. Our data processing pipeline (as seen in Figure~\ref{fig:pipeline}) generates transcripts and NLP tags for each communication. By pre-transcribing with AI tools, we are able to speed up the overall transcription process.\footnote{If you are interested in becoming an annotator, please create an account on the SpokenData transcription platform: \url{http://www.spokendata.com/atco2}.}
The amount of human transcribed~\footnote{This corpus is part of the data released in ELDA catalog: \url{http://catalog.elra.info/en-us/repository/browse/ELRA-S0484/}} data is the package of a four-hour test set, i.e., \textit{ATCO2-test-set-4h corpus}. The data annotators are the final actors involved in the human data annotation step, as shown in Figure~\ref{fig:data_flow}.

\section{Technologies}
\label{sec:technologies-atco2}

In this section, we cover the main tools developed with the ATCO2 corpora. We also list some potential topics that can be explored with it. Moreover, note that the ATCO2 corpora are not limited to the fields covered in this paper, e.g., ASR or NLP, and we hope the community will build on top of it to foster and advance speech and text-based technologies for ATC.

\subsection{Automatic Speech Recognition}
\label{subsec:asr}

One of the principal components of the ATCO2 project is the strong ASR system in order to provide high-quality automatic transcriptions for the collected ATC data. An ASR system is trained to predict the best text translation for the input acoustic signal. Formally speaking, ASR aims to find the best probability candidate output sequence of words from a set of all possible word combinations (or sentences) in a language given a noisy acoustic observation sequence. End-to-end ASR models learn a direct mapping of speech to the output text:

\vspace{-0.5cm}

\begin{equation}
  y^{\ast} = \argmax\limits_{y} \log P(y|x)
\end{equation}

The hybrid (conventional) ASR systems combine three separately trained models: acoustic model (AM), pronunciation model, and language model (LM). The model calculates the conditional probability $p(W|S)$, where $W$ is a sequence of words ($W = \{w_{1}, ..., w_{n}\}$), $S$ is a sequence of input feature vectors representing the acoustic observations ($S = \{s_{1}, ..., s_{t}\}$), and $\mathcal{V}$ is a vocabulary \cite{Jurafsky:2009:SLP:1214993,wang2019overview} (Equation~\ref{eq:asr_pipeline}):

\vspace{-0.5cm}

\begin{align}
  \hat{W} &= \argmax_{W \in \mathcal{V}} p(W|S) \\
          &= \argmax_{W \in \mathcal{V}} p(S|W)p(W) \\
          &= \argmax_{W \in \mathcal{V}} \sum_{P}{p(S|P)}{p(P|W)}{p(W)}
  \label{eq:asr_pipeline}
\end{align}

where $p(S|P)$ is an AM, $p(P|W)$ is a pronunciation model, and $p(W)$ is a LM. One of the advantages of conventional pipeline models is a more transparent optimization of an objective function \cite{kingsbury2009lattice}. Moreover, the LM is trained with unpaired text data and can be easily adapted to a specific domain. This gives conventional models more flexibility and makes them convenient for being used in industrial projects, such as ATC. \\

\subsubsection{Training and Test Data}
\label{subsec:training_data}

\textbf{Training data configuration:} to measure the effectiveness of using automatically transcribed data (ATCO2-PL set) versus using fully-supervised gold annotations, we defined three training scenarios. 

\begin{itemize}
    \item \textbf{\textit{scenario a) only supervised data}}: we employ a mix of public and private supervised ATC databases (recordings with gold annotations). It comprises $\sim$190\,hr of audio data (or 573\,hr after speed perturbation). 
    \item \textbf{\textit{scenario b) only ATCO2-PL 500 hrs dataset}}: we use only a subset of 500\,hr from the ATCO2-PL corpus (see introductory paper~\cite{zuluaga2022atco2}).
    \item \textbf{\textit{scenario c) only ATCO2-PL 2500 hrs dataset}}: same as \textbf{scenario b)}, but instead of only using 500\,hr subset, we use 5 times more, i.e., a 2500\,hr subset. This subset is only used to train a hybrid-based ASR model (CNN-TDNNF) to test the boosting experiments in Section~\ref{subsec:boosting}.
\end{itemize}

\textbf{Test data configuration:} two ATCO2 test sets are used for ASR evaluation, as shown in Table~\ref{tab:asr-results}: ATCO2-test-set-1h (in short ATCO2-1h) and ATCO2-test-set-4h (in short ATCO2-4h). The same test sets are used for boosting experiments presented in Section~\ref{subsec:boosting}.

\subsection{Conventional ASR}
\label{suprebsec:hb-asr}

To obtain automatic transcriptions of the best possible quality for ATCO2 corpus audio, we use a strong hybrid model trained on ATC data only. We train a hybrid-based model for each of the scenarios described above. For \textbf{scenario a)}, an AM was built to include all available 190\,hr-dataset, speech augmentation accounting for 573 hours of data. The model dictionary consists of 30’832 words. For training the acoustic model, we use the Kaldi toolkit~\cite{povey2011kaldi}. The system follows the standard Kaldi recipe, which uses MFCC and i-vectors features. The standard chain training is based on Lattice-free MMI (LF-MMI~\cite{povey2016purely}, which includes 3-fold speed perturbation and one-third frame sub-sampling. The acoustic model is a CNN-TDNNF. The LM is 3-gram trained on the same data as the acoustic model with additional text data coming from additional public resources such as airlines names, airports, ICAO alphabet, and way-points in Europe. 

\textbf{Results \& Analysis:} the results are presented in Table~\ref{tab:asr-results}. We compared three models trained with the same conventional CNN-TDNNF architecture but on different data: scenarios a), b), and c) (see~\ref{subsec:training_data}). The Model a)~in Table~\ref{tab:asr-results} is trained on the "out-of-domain" for ATCO2 but supervised data. The models b) and c) are trained on the "in-domain" ATCO2 data and the difference is only in the size of training set: 500\,h VS 2500\,h. We can see that training on completely unsupervised data can gain good performance. Increasing the size of unsupervised data from 500\,h to 2500\,h, however, does not bring too much improvement: the WER goes from 18.1\% to 17.9\% and from 25.1\% to 24.9\% only for ATCO2-1h and ATCO2-4h, respectively. 

\subsubsection{End-to-End ASR}

Differently from hybrid-based ASR, there exists another paradigm for performing ASR, named end-to-end (E2E) ASR. Here, we aim at directly transcribing speech to text without requiring alignments between input features and output words or characters (i.e., standard procedure in hybrid-based ASR), see Equation~\ref{eq:asr_pipeline}. Recent work on encoder-decoder ASR has shown this step can be removed. E2E can be divided into two categories: Connectionist Temporal Classification (CTC)~\cite{graves2014towards} and attention-based encoder-decoder modeling~\cite{chorowski2015attention}. 
See other works related to encoder-decoder models in~\cite{baevski2020wav2vec,chen2021wavlm,oord2018representation,baevski2019vq,babu2021xls}. 
E2E ASR aims at reducing the expert knowledge need. This makes simpler the overall ASR development, thus, it could have a significant impact in ATC~\cite{zuluaga2022does}. As in this work, we, first of all, address the novelty of the data, its collection and preparation, and do not compare different ASR architectures. Thus, we are leaving the E2E results for other papers.

\begin{table}[t]
    \caption{WER results for {public} ATCO2 test sets with CNN-TDNNf models trained on different data (scenarios (a-c)).
    }
    \centering
    \label{tab:asr-results}
    \begin{tabular}{l | cc}
        \toprule
        \cellcolor{Gray}\textbf{Model} & \multicolumn{2}{c}{\cellcolor{Gray} \textbf{Test sets}} \\
        \cmidrule(lr){1-3}         
        \cellcolor{Gray}\textbf{CNN-TDNNF} &
        \textbf{ATCO2-1h }& \textbf{ATCO2-4h} \\
        \midrule
        \textbf{scenario a) only supervised 573\,h dataset} & 24.5 & 32.5 \\
        \textbf{scenario b) only ATCO2-PL 500\,h dataset} & 18.1 & 25.1 \\
        \textbf{scenario c) only ATCO2-PL 2500\,h dataset} & 17.9 & 24.9 \\
        \bottomrule
        \end{tabular}
\end{table}

\subsubsection{Callsign Boosting}
\label{subsec:boosting}

To further improve the prediction made by an ASR system, along with speech input one can use other information available from context. For the ATC domain, such context information may be the data received from radar. Every moment radar registers aircraft that are currently in the airspace listing unique identifiers of those aircraft, i.e., `callsigns'. With the radar data, we know exactly what callsigns are especially probable to appear in the conversation. This knowledge allows us to bias the system outputs towards these registered callsigns and to increase the probability that they are recognized correctly. A callsign is typically a sequence of an airline name, letters, and digits, which in speech turns into a sequence of words. In ASR, the target sequences of words can be boosted during WFST (Weighted Finite State Transducer) decoding by adjusting the weights in the prediction graphs, called `lattices'~\cite{hall2015composition,aleksic2015bringing,serrino2019contextual,nigmatulina2022two}. The rescoring of lattices is done with the Finite State Transducer (FST) operation of composition between lattices produced by an ASR system and an FST created with the target word sequences and discount weights (Equation~\ref{eq:fst-biased-composition}):
\vspace{-0.1cm}

\begin{equation}
  biased\_Lattices = Lattices \circ biasing\_FST
  \label{eq:fst-biased-composition}
\end{equation}

Biasing the lattice toward the context callsigns usually allows us to considerably improve their recognition in the final outputs (Table~\ref{tab:atco2-boosting}). The results of different experiments on the ATC data proved that applying the lattice rescoring method on top of ASR predictions leads to higher accuracy of automatic transcriptions, first of all, callsigns \cite{zuluagagomez21_interspeech,nigmatulina2022two}. Therefore, lattice rescoring was used for all transcriptions of the ATCO2 data. \\

\begin{table}[t]
    \caption{Results for boosting experiment on ATCO2 corpora. Results are listed for the CNN-TDNNf model trained with either all supervised data or 500\,hrs or 2500\,hrs of ATCO2 corpora. The top results per block are \textbf{highlighted in bold}. The best result per column is marked with \underline{underline}. $^\dagger$1h public test set. $^\ddagger$4h full test set. Results are obtained with offline CPU decoding. $^\mathparagraph$word error rates only on the sequence of words that compose the callsign in the utterance.}
    \centering
    \label{tab:atco2-boosting}
    \begin{tabular}{l |ccc|ccc}
        \toprule
        \rowcolor{Gray} \multicolumn{1}{c}{\textbf{Boosting}} & \multicolumn{3}{c}{\textbf{ATCO2-test-set-1h$^\dagger$}} & \multicolumn{3}{c}{\textbf{ATCO2-test-set-4h$^\ddagger$}} \\
        \cmidrule(lr){2-4} \cmidrule(lr){5-7}
         & WER & EntWER$^\mathparagraph$ & ACC & WER & EntWER$^\mathparagraph$& ACC \\
        \midrule
        & \multicolumn{6}{l}{\cellcolor{Gray} \textbf{scenario a) only supervised dataset}} \\        
        \midrule
        Baseline & 24.5 & 26.9 & 61.3 & 32.5 & 36.7 & 42.4 \\
        Unigrams & 24.4 & 25.5 & 63.2 & 33.1 & 35.0 & 45.8 \\
        N-grams & 23.8 & 23.8 & 66.4 & 31.3 & 33.7 & 47.9 \\
        GT boosted & \textbf{22.9} & \textbf{19.1} & \textbf{75.2} & \textbf{29.7} & \textbf{29.1} & \textbf{58.5} \\
        \midrule
        & \multicolumn{6}{l}{\cellcolor{Gray} \textbf{ scenario b) only ATCO2-PL 500\,hrs dataset}} \\        
        \midrule  
        Baseline & 18.1 & 16.2 & 71.2 & 25.1 & 24.8 & 62.6 \\
        Unigrams & 19.1 & 14.6 & 74.2 & 26.0 & 22.8 & 65.6 \\
        N-grams & 17.5 & 13.5 & 75.3 & 24.3 & 21.4 & 66.6 \\
        GT boosted & \textbf{16.3} & \textbf{6.9} & \textbf{88.9} & \textbf{22.5} & \textbf{13.0} & \textbf{82.9} \\
        \midrule
        & \multicolumn{6}{l}{\cellcolor{Gray} \textbf{ scenario c) only ATCO2-PL 2500\,hrs dataset}} \\        
        \midrule  
        Baseline & 17.9 & 16.7 & 70.5 & 24.9 & 24.2 & 62.0 \\
        Unigrams & 18.3 & 14.4 & 73.8 & 25.6 & 22.0 & 65.9 \\
        N-grams & 17.3 & 14.2 & 74.3 & 24.3 & 21.1 & 66.5 \\
        GT boosted & \underline{\textbf{15.9}} & \underline{\textbf{6.5}} & \underline{\textbf{89.4}} & \underline{\textbf{22.2}} & \underline{\textbf{12.5}} & \underline{\textbf{83.9}} \\
        \bottomrule
        \end{tabular}
\end{table}

\textbf{Results \& Analysis:} in Table~\ref{tab:atco2-boosting}, we report the results for the out-of-domain (ATC supervised) and in-domain (ATCO2-500h/2500hr) ATC models. Both acoustic models are trained with CNN-TDNNF architecture following the standard Kaldi recipe, as described in~\ref{suprebsec:hb-asr}.

Biasing n-grams, compared to biasing only a ground truth (GT) callsign, can be used in a real-life scenario and with the real-time ASR: a new contextual FST is generated on-the-fly every time when new data comes from radar. The results of biasing a GT callsign are given here to illustrate the oracle performance of the biasing method. Overall, decoding with n-grams biasing always helps to achieve better performance, especially for callsigns, with a relative improvement of 15.0\% and 12.8\% for callsign recognition and of 3.4\% and 2.4\% for the entire utterance on ATCO2-test-set-1h and ATCO2-test-set-4h test sets, respectively.

The size of biasing FST depends on the number of entities to boost. Too many entities may decrease the effectiveness of the biasing method, as the more non-true entities are boosted the less the correct sequence is prominent. The previous results show that the optimal size of biasing FST highly depends on the data but generally, the performance begins to degrade when the number of contextual entities exceeds 1000~\cite{chen2019end}. For our experiments, we have on average 214 biasing entities per utterance in the ATCO2-test-set-4h and 140 biasing entities per utterance in the ATCO2-test-set-1h corpus.

\subsection{Natural Language Understanding of Air Traffic Control Communications}

\begin{figure}[t]
    \centering
    \includegraphics[width=0.9\linewidth]{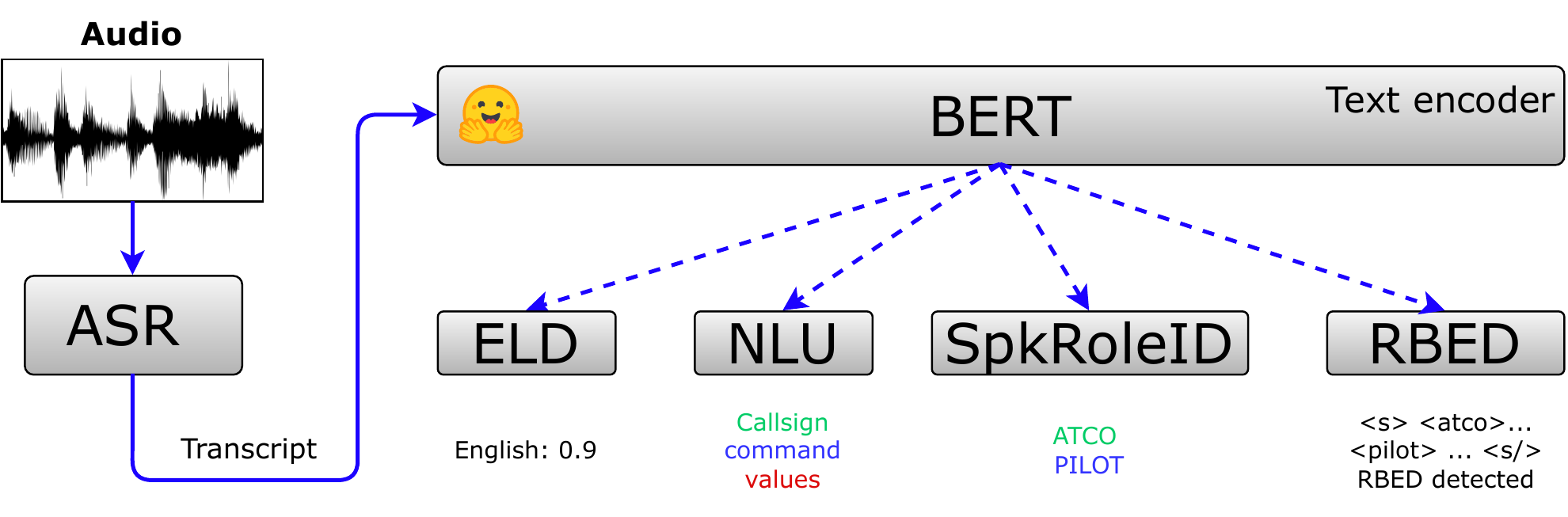}
    \caption{\textit{Main automatic speech recognition and understanding tasks that can be achieved with the ATCO2 corpora}. ELD: English Language Detection; NLU: natural language understanding, e.g, callsign highlighting; SPKRoleID: speaker role identification; RBED: read-back error detection.}
    \label{fig:ASRU}
\end{figure}

Natural language understanding (NLU) is a subfield of NLP that focuses on the ability to understand and interpret human language. NLU involves the development of algorithms and models that can extract meaning and intent from text and/or spoken communication. NLU involves several subtasks, including (i)~Named Entity Recognition~\cite{yadav2018survey,sharma2022named}, which aims at identifying entities in text, such as people, places, and organizations. (ii)~Part-of-Speech Tagging (POS), identifying the grammatical role of each word in a sentence~\cite{collobert2011natural}, similar to sequence classification (see Section~\ref{subsub:text-spkid}). (iii)~Sentiment Analysis, identifying the emotional tone of a piece of text~\cite{birjali2021comprehensive}. (iv)~Relationship Extraction, identifying the relationships between entities in text. (v)~Question Answering, understanding, and answering natural language questions. The following subsections cover each of the proposed NLU submodules that can be developed with ATCO2 corpora, like the ones presented in Figure~\ref{fig:ASRU}.

\subsubsection{Named Entity Recognition for Air Traffic Control Communications}
\label{subsec:ner-system}

In ATC communications, NLU can be used to automatically analyze and interpret the meaning of spoken messages between pilots and ATCos. This can help improve communication accuracy and efficiency, as well as assist in identifying emergency situations and other critical events. Additionally, NLU can help to extract important information, such as flight numbers, callsigns, or airport codes, which in turn can aid ATCos to manage traffic more efficiently. Overall, the use of NLU in ATC helps improve communication accuracy and efficiency, enhances safety, and provides valuable data for analysis and decision-making. 

\begin{figure}[t]
    \centering
    \includegraphics[width=0.95\linewidth]{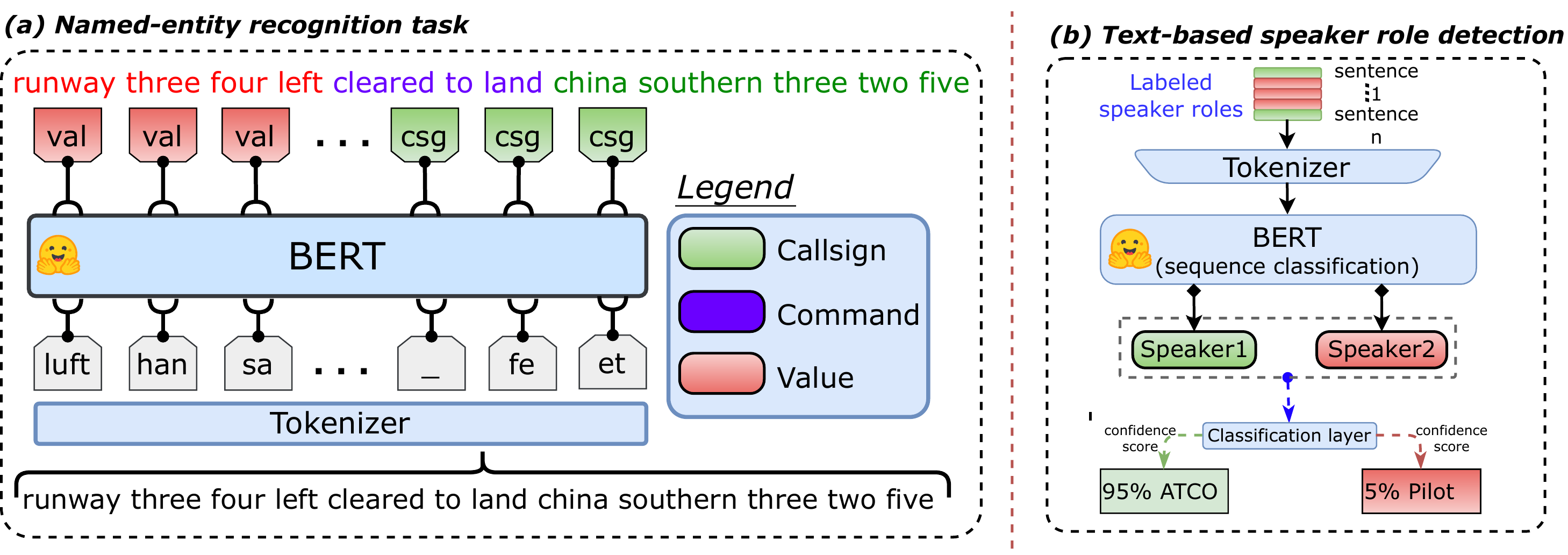}
    \caption{\textit{(a) Named entity recognition and (b) speaker role detection module based on sequence classification (SC).} Both systems are based on fine-tuning a pre-trained BERT~\cite{devlin2018bert} model on ATC data. The NER systems recognize callsign, command, and values, while the SC assigns a speaker role to the input sequence. Taken from~\cite{zuluaga2022atco2}.}
    \label{fig:ner-module}
\end{figure}

In this work, one of the main tasks is to understand and extract high-level information within ATC communication. Therefore, we develop a NER system tasked to extract this information, as depicted in Figure~\ref{fig:ner-module} (a). For instance, consider the following transcribed communication (taken from Figure~\ref{fig:atco2-ecosystem}):

\vspace{0.2cm}

\noindent \textbf{ASR transcript:} \textcolor{black}{\dashuline{runway three four left cleared to land china southern three two five}}, 

\vspace{0.2cm} 

\noindent would be converted to high-level entity format with the NER system to:

\vspace{0.2cm} 

\noindent \textbf{Output:}
\textcolor{red}{<value> \uwave{runway three four left} </value>}
\textcolor{purple}{<command> \dotuline{cleared to land} </command>}
\textcolor{teal}{<callsign> \dashuline{china southern three two five} </callsign>} . 
 
\vspace{0.2cm}

In this work, we developed two systems based on Transformers~\cite{vaswani2017attention} to extract and tag this information from ATC communications, i.e., a pre-trained BERT~\cite{devlin2018bert} model and RoBERTa~\cite{liu2019roberta} model. \\

\textbf{Experimental setup:} we fine-tune a pre-trained BERT and RoBERTa\footnote{The pre-trained version of \texttt{BERT-base-uncased}~\cite{devlin2018bert} with 110M parameters, URL: \url{https://huggingface.co/bert-base-uncased}. The pre-trained version of \texttt{RoBERTa-base}~\cite{liu2019roberta} with 123M parameters, URL: \url{https://huggingface.co/roberta-base}} model on the NER task (see Figure~\ref{fig:ASRU}). We download the pre-trained models from HuggingFace~\cite{wolf2020transformers,lhoest2021datasets}. For training, we use the full ATCO2-test-set-4h, which contains $\sim$3k sentences. In this dataset, each word is annotated together with a predefined class, as follows:
\textcolor{teal}{\dashuline{callsign}}, 
\textcolor{purple}{\dotuline{command}}, 
\textcolor{red}{\uwave{values}}, and
\textcolor{black}{\uwave{UNK}} (everything else). In order to fine-tune the model, we append a layer on top of the BERT model by a feedforward network with a dimension of 8.\footnote{We define two outputs per class. See the class structures in Section 3.3 of~\cite{zuluaga2021bertraffic} and in~\cite{nigmatulina2022two}.}. Due to the lack of gold annotations, we perform a 5-fold cross-validation scheme to avoid overfitting. The reader interested in developing their own NER system for ATC is redirected to the open-source GitHub repository of the ATCO2 corpus.\footnote{GitHub repository: \url{https://github.com/idiap/atco2-corpus}} We fine-tune each model on an NVIDIA GeForce RTX 3090 for $\sim$10k steps. During experimentation, we use a linear learning rate scheduler with an initial learning rate of $\gamma=5\mathrm{e}{-5}$, dropout~\cite{srivastava2014dropout} of $dp=0.1$, and GELU activation function~\cite{hendrycks2016gaussian}. We also employ gradient norm clipping~\cite{pascanu2013difficulty} for regularization and AdamW as optimizer~\cite{loshchilov2018decoupled}. Each model during the cross-validation scheme uses an effective batch size of 32. \\

\textbf{Evaluation metric:} we evaluate both BERT and RoBERTa NER systems with a binary classification metric named, F-score. Particularly, the F1-score, defined in Equation~\ref{eq:f1-score}, represents the harmonic mean of precision and recall. Recall, as defined in Equation~\ref{eq:recall}, is the ratio of $TP$ to all samples that should have been identified as positive (including false negatives ($FN$)). Precision, as described in Equation~\ref{eq:precision}, is the ratio of true positive ($TP$) results to all positive results (including false positives ($FP$)):

\begin{equation}
    \label{eq:precision}
    Precision = \frac{TP}{TP+FP}
\end{equation}

\begin{equation}
    \label{eq:recall}
    Recall = \frac{TP}{TP+FN}
\end{equation}

\begin{equation}
    \label{eq:f1-score}
    F1 = \frac{2*Precision*Recall}{Precision+Recall} = \frac{2*TP}{2*TP+FP+FN}
\end{equation}

\vspace{0.2cm}

\textbf{Results \& Analysis:} the NER system's performance is evaluated on the ATCO2-test-set-4h corpus using a 5-fold cross-validation scheme, with five fine-tuning runs using different training seeds. Table~\ref{tab:ner_results} presents the performance metrics for callsign, command, and values classes of two Transformer-based~\cite{vaswani2017attention} models, namely BERT-base and RoBERTa-base. Although both models achieve similar F1-scores, we provide analysis for the BERT-based NER system, which achieves an F1-score of over 97\% for the \textcolor{teal}{\dashuline{callsign}} class, while the \textcolor{purple}{\dotuline{command}} and 
\textcolor{red}{\uwave{values}} classes lag behind with F1-scores of 81.9\% and 87.1\%, respectively. We hypothesize that the command class contains higher complexity when compared to the other two classes, values and callsigns. Values are mostly composed of defined keywords (e.g., flight level) followed by cardinal numbers (e.g., “one hundred”), while callsigns follow a well-defined structure of airline designators and a set of numbers or letters spoken in ICAO format~\cite{allclear}. These characteristics make it easier for the NER system to correctly detect them.

One potential method for increasing the performance of the NER system for the command and values classes is to incorporate plausible commands and values in real-time, depending on the situation of the surveillance data. This can be achieved using the boosting technique, as described in Section~\ref{subsec:boosting}. Although the results with boosting callsigns are reported in Table~\ref{tab:atco2-boosting}, further investigation is needed to assess the impact of boosting on the command and values classes.

\begin{table}[t]
    \caption{Different performance metrics for callsign, command, and values classes of the NER system. Results are averaged over a 5-fold cross-validation scheme on \textit{ATCO2-test-set-4h corpus} in order to mitigate overfitting. We run five times fine-tuning with different training seeds (2222/3333/4444/5555/6666). Results are reported on two Transformer-based models. @P, @R and @F1, refers to precision, recall and F1-score, respectively. }
    \centering
    \label{tab:ner_results}
    \begin{tabular}{l ccc ccc ccc}
        \toprule
        Model & \multicolumn{3}{c}{\textbf{\textcolor{teal}{Callsign}}} & \multicolumn{3}{c}{\textbf{\textcolor{purple}{Command}}} & \multicolumn{3}{c}{\textbf{\textcolor{red}{Values}}} \\
        \cmidrule(lr){2-4}
        \cmidrule(lr){5-7}
        \cmidrule(lr){8-10}
        & @P & @R & @F1 & @P & @R & @F1 & @P & @R & @F1 \\
        \midrule
        Bert-base & 97.1 & 97.8 & 97.5 & 80.4 & 83.6 & 82.0 & 86.3 & 88.1 & 87.2 \\
        RoBERTa-base & 97.1 & 97.7 & 97.5 & 80.2 & 83.7 & 81.9 & 85.6 & 88.6 & 87.1 \\
        \bottomrule
    \end{tabular}
\end{table}

\subsubsection{Text-based Speaker Role Detection}
\label{subsub:text-spkid}

Sequence classification (SC) is a type of machine learning task that involves assigning a label or a category to a sequence of data points~\cite{he2020survey,zhou2015pattern}. The data points in the sequence can be of various types, such as text, audio, or numerical data, and the label assigned to the sequence can also be of different types, such as binary (e.g., positive or negative sentiment~\cite{birjali2021comprehensive}) or multiclass. Sequence classification can also be used to automatically classify ATC communication sequences into various categories. This technique can be applied to both audio and text data, making it a versatile tool to provide a high-level understanding of the communication at hand. 

In scenarios where the PTT signal is unavailable and there exists only a monaural communication channel, it can be challenging to discern the identity of a speaker. Therefore, it becomes crucial to distinguish between the ATCo and the pilot over the target communications. As a potential solution, we propose an alternative approach that utilizes a speaker role detection (SRD) system based on SC. The system gets text as an input, and it returns as output a category where the communication falls, either uttered by the ATCo or the pilot. 
In recent years, there has been a growing interest in using deep learning techniques, such as the Transformer-based models~\cite{vaswani2017attention}, to improve the performance of SC for SRD in ATC communications. Here, we ablate three types of such models, (i)~BERT~\cite{devlin2018bert}, (ii)~RoBERTa~\cite{liu2019roberta}, and (iii)~DEBERTA~\cite{he2021debertav3}. These models have been shown to achieve state-of-the-art performance on a wide range of sequence classification tasks, including SRD for ATC. The proposed SRD is illustrated in Figure~\ref{fig:ner-module} (b).

Overall, the SRD and speaker diarization (see Section~\ref{subsub:text-dia}) tasks can leverage the fact that ATC dialogues follow a well-defined lexicon and dictionary with simple grammar. This standard phraseology has been defined by the ICAO~\cite{allclear} to guarantee safety and reduce miscommunications between the ATCos and pilots. Therefore, previous work has shown the potential in performing SRD in an E2E manner on the text-level, as presented here, see in~\cite{zuluaga2021bertraffic,prasad2022grammar,zuluaga2023avirtual}. \\

\textbf{Experimental setup:} the SRD system is built on top of pre-trained models (BERT~\cite{devlin2018bert}, RoBERTa~\cite{liu2019roberta}, and DEBERTA~\cite{he2021debertav3}), which are downloaded from HuggingFace~\cite{wolf2020transformers,lhoest2021datasets}. Here, the experimental setup is exactly the same as the one described for the NER system, including the training hyperparameters. For further details, we redirect the reader to Section~\ref{subsec:ner-system}. Still, the SRD model is fine-tuned on the SC rather than on the NER task. Further, we define an output layer with 2 units (classes), one for ATCo and one for pilot. \\

\begin{table}[t]
    \caption{\textit{ATCO}/\textit{PILOT} F1-scores for speaker role identification on ATCO2-test-set-4 test set. Metrics reported with three different Transformer-based models (BERT~\cite{devlin2018bert}, RoBERTa~\cite{liu2019roberta}, deBERTa-V3~\cite{he2021debertav3}). All models are the 'base' version, e.g., \texttt{bert-base}. Numbers in \textbf{bold} refer to the top performance per split, i.e., \textbf{\textcolor{teal}{ATCO}} or \textbf{\textcolor{magenta}{PILOT}}. Results are averaged over a 5-fold cross-validation scheme on \textit{ATCO2-test-set-4h corpus} in order to mitigate overfitting. Each round of fine-tuning is run five times with different training seeds (2222/3333/4444/5555/6666).}
    \centering
    \label{tab:spkid-results}
    \begin{tabular}{l cccccc}
        \toprule
        \rowcolor{Gray} \textbf{Training Corpora} & \multicolumn{2}{c}{\textbf{BERT}} & \multicolumn{2}{c}{\textbf{DEBERTA}} & \multicolumn{2}{c}{\textbf{ROBERTA}} \\
        \cmidrule(lr){2-3}
        \cmidrule(lr){4-5}
        \cmidrule(lr){6-7}
        & \textbf{\textcolor{teal}{ATCO}} & \textbf{\textcolor{magenta}{PILOT}} & \textbf{\textcolor{teal}{ATCO}} & \textbf{\textcolor{magenta}{PILOT}} & \textbf{\textcolor{teal}{ATCO}} & \textbf{\textcolor{magenta}{PILOT}} \\
        \midrule
        LDC-ATCC & 82.4 & 79.2 & 82.4 & 79.6 & 84.0 & 80.2 \\
        UWB-ATCC & 86.2 & 83.2 & 86.8 & 84.0 & 87.0 & 82.8 \\
        \midrule
        $\hookrightarrow$ + LDC-ATCC &\textbf{ 87.6} & \textbf{85.2} & \textbf{88.8} & \textbf{85.8} & \textbf{88.0} & \textbf{84.2} \\
        \bottomrule
    \end{tabular}
\end{table}

\textbf{Results \& Analysis:} we evaluate the SRD system on ATCO2-test-set-4h corpus. Differently from the NER system, here, we have access to two training corpora,  
LDC-ATCC\footnote{The Air Traffic Control Corpus (LDC-ATCC) corpus, see URL: \url{https://catalog.ldc.upenn.edu/LDC94S14A}. It consists of audio recordings in the area of ASR for air traffic control communications. We use the metadata along the transcrips to perform research on NLU for ATC, i.e., speaker role detection. The data files are sampled at 8 kHz, 16-bit linear, with continuous monitoring and without squelch or silence elimination.} and UWB-ATCC\footnote{The UWB-ATCC corpus by the University of West Bohemia can be downloaded for free in the following URL: \url{https://lindat.mff.cuni.cz/repository/xmlui/handle/11858/00-097C-0000-0001-CCA1-0}. UWB-ATCC contains recordings of air traffic control communication. The speech is manually transcribed and labeled with the speaker information, thus it can be used for speaker role detection} datasets. We evaluate the SRD under two considerations: (i)~ablations of different pre-trained models for SRD on ATC communications and (ii)~low-resource and incremental training scenario.

\textbf{\textit{(i) Analysis of the Impact of Pre-trained Models and Training Data Type}} \quad In this scenario, we evaluate the impact of pre-trained models and training data on the SRD task for ATC data. To this end, we compare the performance of three Transformer-based~\cite{vaswani2017attention} models, including BERT, RoBERTa, and deBERTa-V3, trained on two different corpora, LDC-ATCC and UWB-ATCC, and evaluate them on the ATCO2-test-set-4h corpus. The F1-scores for SRD are reported separately for ATCo and pilot speakers in Table~\ref{tab:spkid-results}. Our results show that all the models achieved comparable F1-scores, ranging from 87-88\% for ATCo and 84-85\% for pilots. These findings suggest that the SRD task for ATC data is not significantly sensitive to the choice of pre-trained models. However, we observe that models trained on UWB-ATCC outperform those trained on LDC-ATCC, with up to 4\% absolute improvement in F1-scores. For instance, \textit{\textbf{BERT-model}} with LDC-ATCC $\rightarrow$ UWB-ATCC comparison: 82.4\% $\rightarrow$ 86.2\% for ATCo and 79.2\% $\rightarrow$ 83.2\%, for Pilot. Additionally, we find that combining both datasets leads to a 1\% absolute improvement in F1-scores. Overall, our study highlights the importance of selecting appropriate training data for the SRD task in ATC data and suggests that using multiple datasets can lead to improved performance. The findings also suggest that the choice of pre-trained models has a relatively minor impact on the SRD task for ATC data.

\begin{figure}[t]
    \centering
    \includegraphics[width=0.9\linewidth]{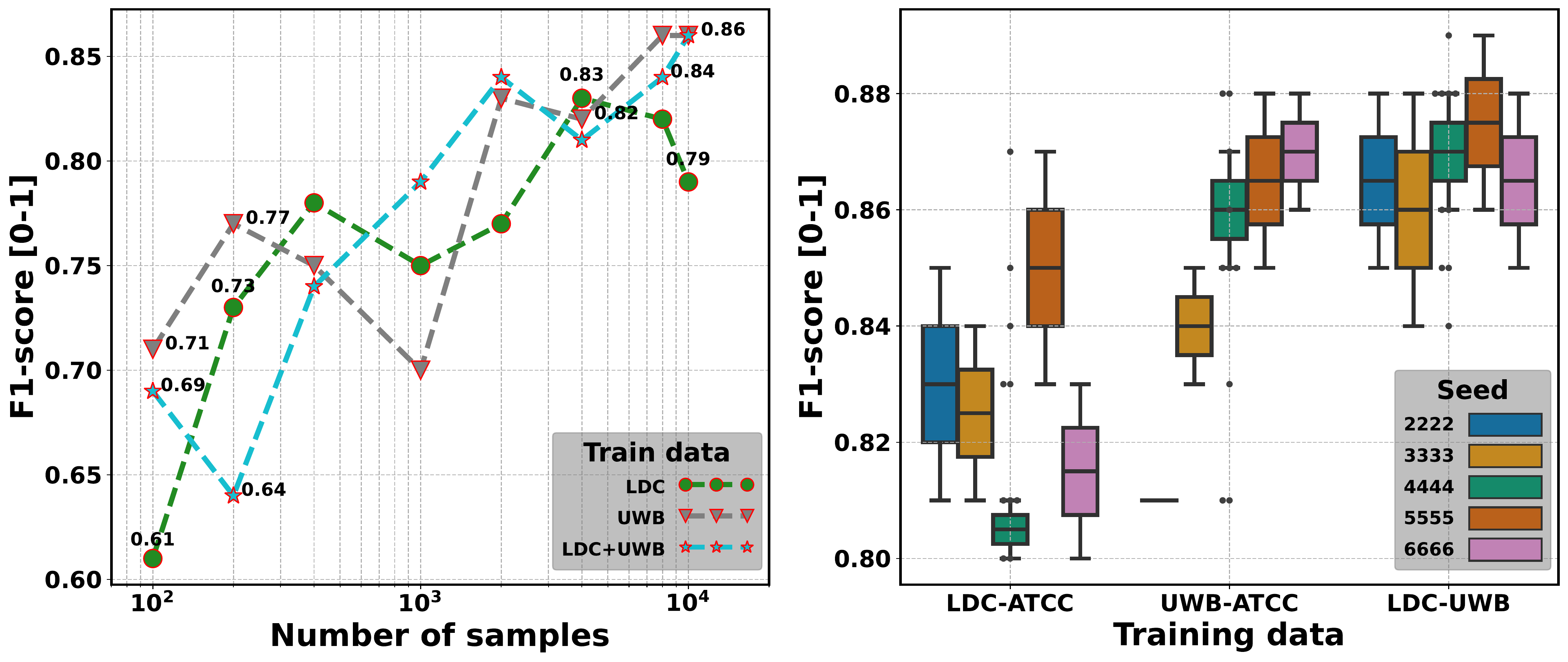}
    \caption{\textit{Metrics for the speaker role detection system (introduced in~\cite{zuluaga2022atco2})}. Metrics reported only on \textit{ATCO2-test-set-1h corpus} with a \texttt{bert-base-uncased} model trained with different datasets from Table~\ref{tab:databases}. Left plot: ablation of the F1-score versus the number of samples used to train the system. Right plot: F1-score for models trained with different training seeds. The box plot depicts the performance variability when splitting the test set into ATCo and pilot subsets.}
    \label{fig:results-spkID}
\end{figure}

\textbf{\textit{(ii) Analysis of the Impact of Data Quantity on Speaker Role Detection}} \quad In this study, we aim to evaluate the impact of the number of text samples on the performance of SRD. The results of this analysis are illustrated in the left panel of Figure~\ref{fig:results-spkID}, where the F1-score on the ATCO2-test-set-4h is plotted against the number of samples in a logarithmic scale on the x-axis. Interestingly, we found that as few as 100 samples are necessary to achieve a reasonably good F1-score of 60\% on SRD. Notably, the UWB-ATCC appears to be more informative for the BERT model, which achieves an F1-score of 71\% with only 100 training samples. Increasing the training data to 1000 samples further improves the performance, resulting in F1-scores near 80\% (LDC-ATCC + UWB-ATCC). These findings are significant, considering that the annotation of ATC communications is generally expensive and time-consuming. In the right panel of Figure~\ref{fig:results-spkID}, we present a box plot that shows the variation of the BERT model's performance when fine-tuned on SRD with different training seeds. Each box represents the variation of the model between the ATCo and pilot subsets, over the 5-fold cross-validation scheme. Overall, the results indicate that increasing the training data leads to better performance and more consistent results. These observations highlight the importance of selecting a suitable training set size for speaker role detection tasks.

\subsubsection{Text-based Diarization}
\label{subsub:text-dia}

In addition to only detecting roles in a given ATC communication (e.g., SRD), there are cases where multiple segments end up in the same recording/communication. The task that solves this issue is known as, speaker diarization (SD). SD answers the question \textit{``who spoke when?''}. Here, the system receives an audio signal or recording (or text in our case) and detects the speaker changes or segmentation and the speaker role. The main parts of an SD system are (i)~segmentation, (ii)~embedding extraction, (iii)~clustering, and (iv)~labeling (similar to SRD). SD is normally performed on the acoustic level, previous work based on mel filterbank slope and linear filterbank slope was covered in~\cite{madikeri2014filterbank}. Speaker discriminative embeddings such as x-vectors are investigated in~\cite{sell2018diarization}, and more recently a variational Bayesian hidden Markov model (VBx) was investigated in~\cite{valente2010variational,landini2022bayesian}, which is the SD system used during the data collection stage of ATCO2, see Section~\ref{subsec:data-collection-pipeline}. State-of-the-art SD systems are based on the E2E paradigm, named E2E neural diarization (EEND). This approach was introduced in~\cite{fujita2019end_free,fujita2019end}. Here, an SD model is trained jointly to perform extraction and clustering. Here, differently from SRD, we only used the BERT~\cite{devlin2018bert} pre-trained model. \\

\textbf{Experimental setup:} the SD system is built on top of a pre-trained BERT model downloaded from HuggingFace~\cite{wolf2020transformers,lhoest2021datasets}. As in the NER and SRD system, here the experimental setup is the same, this also includes the training hyperparameters.\footnote{See the open-source GitHub repository of ATCO2: \url{https://github.com/idiap/atco2-corpus}}  For further details we redirect the reader to Section~\ref{subsec:ner-system}. The SD model is fine-tuned on the NER task, where each speaker role (ATCo or pilot) is a class. Therefore, we have 2 tags per class, accounting for four classes in total. Readers are directed to our paper on text-based SD presented at The 2022 IEEE Spoken Language Technology Workshop (SLT 2022), see~\cite{zuluaga2021bertraffic}. \\

\textbf{Evaluation metric:} to score the text-based SD system, we use Jaccard Error Rate (JER) metric. JER is a recent metric introduced in~\cite{ryant19_interspeech} that aligns with speaker diarization. JER aims at avoiding the bias that the predominant speaker might cause. I.e., JER evaluates equally all speakers. The JER is defined in Equation~\ref{eq:jer}:

\begin{equation}
    \label{eq:jer}
    JER=1-\frac{1}{\text{\#speakers}}\sum_{\text{speaker}}\text{max}_{\text{cluster}}\frac{|\text{speaker}\cap\text{cluster}|}{|\text{speaker}\cup\text{cluster}|},
\end{equation}

\noindent where (i) $\text{speaker}$ is the selected speaker from reference and (ii) \textit{$\text{max}_{\text{cluster}}$} is the cluster from the system with maximum overlap duration with the currently selected speaker. \\

\textbf{Results \& Analysis:} we evaluate the SRD system on ATCO2-test-set-4h corpus. Differently from the NER system but similar to SRD, here, we have access to two training corpora: LDC-ATCC and UWB-ATCC datasets. We evaluate the SD under one consideration: (i) low-resource and incremental training scenario.

\begin{figure}[t]
    \centering
    \includegraphics[width=0.9\linewidth]{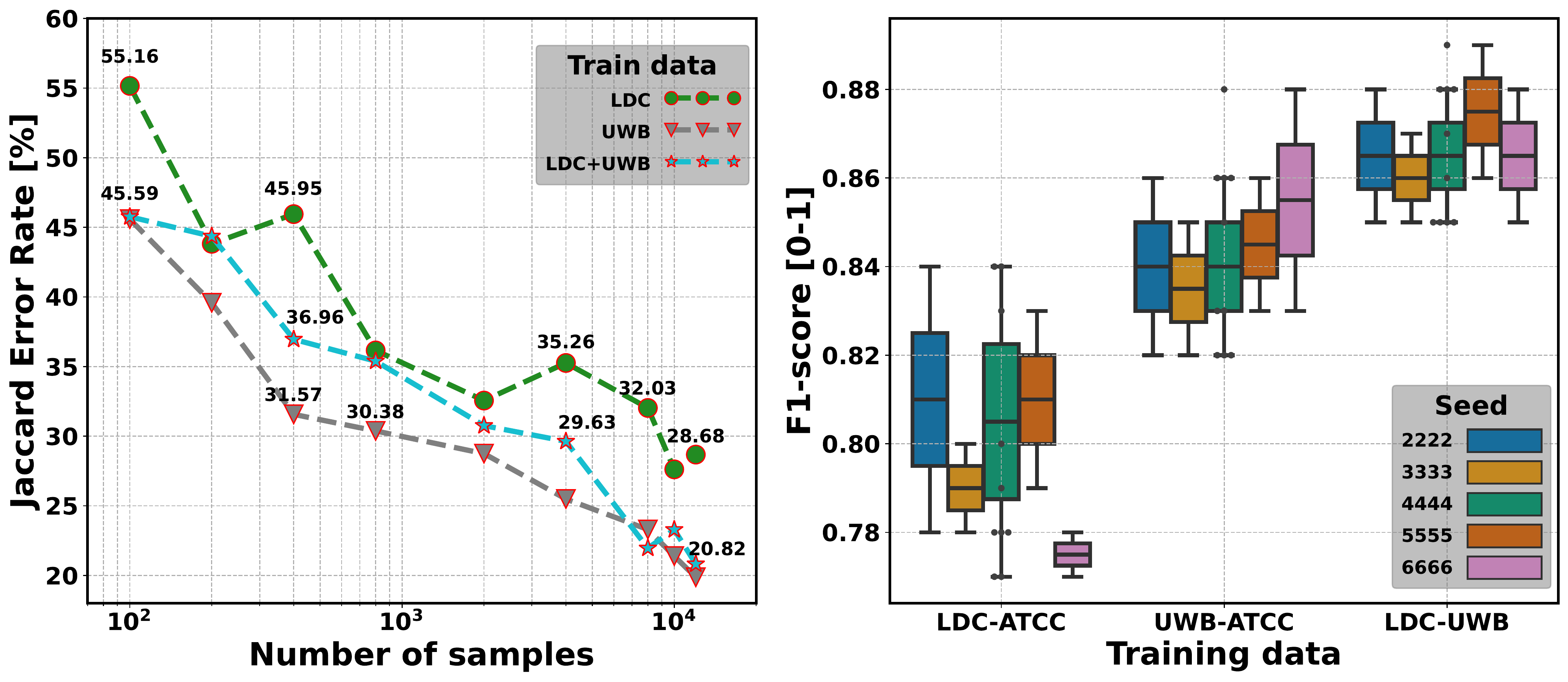}
    \caption{\textit{Metrics for the text-based diarization system (introduced in~\cite{zuluaga2021bertraffic,zuluaga2022atco2}).} Metrics reported only on \textit{ATCO2-test-set-4h corpus} with a \texttt{bert-base-uncased} model trained with different datasets from Table~\ref{tab:databases}. Left plot: ablation of the Jaccard Error Rate versus the number of samples used to train the system. Right plot: F1-score for models trained with different training seeds. The box plot depicts the performance variability when splitting the test set by ATCo and pilot subsets.}
    \label{fig:results-text-dia}
\end{figure}

\textbf{\textit{(i) Analysis of the Impact of Data Quantity on Text-based Speaker Diarization}} \quad In this study, we aim to evaluate the impact of the number of text samples on the performance of SD. The results of this analysis are illustrated in the left panel of Figure~\ref{fig:results-text-dia}, where the JER (the lower the better) on the ATCO2-test-set-4h is plotted against the number of samples in a logarithmic scale on the x-axis. We found that as few as 100 samples are necessary to achieve a JER score of 45.6\% (LDC-UWB). Similar to SRD, the UWB-ATCC dataset seems to be more informative in the SD system. For instance, under the 1000 samples scenario, 
we noted a 5\% absolute JER reduction if UWB-ATCC is used. Furthermore, increasing the training data to 10k samples improved the performance, resulting in JER scores near to 20\% (LDC+UWB). A more appropriate comparison of text and acoustic-based SD for ATC communications can be found in our previous work~\cite{zuluaga2021bertraffic}. Additionally, in the right panel of Figure~\ref{fig:results-spkID}, we present a box plot that shows the variation of the BERT-based SD model's performance when fine-tuned with different training seeds. Each box represents the variation of the model between the two proposed classes: ATCo and pilot, over the 5-fold cross-validation scheme. The results are listed with F1-scores. Overall, we can conclude that the UWB-ATCC dataset is more informative for the SD model in comparison to LDC-ATCC dataset.

\subsubsection{End of Communication Detection}

In the ATC domain, it is crucial to detect the end of communications. While PTT signals are commonly available, there are cases where they are not, and in such scenarios, ATCO2 corpora can be leveraged to develop end-of-communication detection systems using either acoustic or text-based approaches. Acoustic-based systems, known as VAD, perform their task prior to the ATC communication being sent to the ASR system~\cite{sarfjoo2020speech,ariav2019end}, but may require the integration of a new independent module into the recognition pipeline. Text-based systems rely on strong artificial intelligence models like BERT, and previous studies in ATC~\cite{zuluaga2021bertraffic} have shown their effectiveness in detecting callsigns~\cite{lin2021spoken}, commands~\cite{zuluaga2022atco2}, and end-of-communication signals from transcripts generated by an ASR system.

\subsubsection{Read-Back Error Detection}

Read-back error detection (RBED) is a crucial responsibility of ATCos as undetected errors can lead to dangerous situations in ATM. Despite the infrequency of communication errors in air traffic control (ATC), they still have the potential to cause significant safety issues, with some transmissions containing multiple errors. Previous research has shown that between every hundredth~\cite{cardosi1993analysis,cardosi1994analysis} to every sixteenth ATC communication may contain an error~\cite{prinzo2008computation}. Luckily, the capability of en-route ATCos to detect such errors is quite high, with studies indicating that they are able to detect up to 90\% of pilot read-back errors~\cite{cardosi1993analysis}. Although, in general, read-back errors are quite rare, preventing even one incident due to automatic RBED can make an important difference in ensuring ATM safety. To support ATCos in this task, previous projects employ ASRU engines to extract high-level information from ATC communications~\cite{helmke2018ontology}. 
In~\cite{hartmut2022readback} are proposed two approaches for performing RBED. One system is based on rules, while a second system is a data-driven sequence classifier based on BERT-like pre-trained encoder, named RoBERTa~\cite{liu2019roberta}. Here, the input sequence is a concatenation of ATCo and pilot utterance transcriptions with a special separator token \texttt{[SEP]} between them. They show that combining these approaches results in an 81\% RBED rate in real-life voice recordings from Isavia's en-route airspace. They also cover a proof-of-concept trial with six ATCos producing challenging, artificial read-back error samples. 

Nevertheless, a main issue with well-known past projects, such as HAAWAII or MALORCA, is that their data cannot be publicly shared. In contrast, ATCO2 corpora are open to the public, e.g., \textit{ATCO2-test-set-1h} set can be accessed for free and practitioners can follow previous research to implement an RBED module. 

\subsubsection{English Language Detection}

Currently, we have developed and deployed a suitable English language detection system (ELD) to discard non-English utterances in newly collected data. We tested a state-of-the-art acoustic-based system with an x-vector extractor. We also came up with the idea of using an NLP approach that processes ASR output with word confidence for the ELD. Finally, our experiments show that the ELD based on NLP is superior to the acoustic approach in both detection accuracy and computational resources. Moreover, the NLP approach can use outputs from several ASR systems jointly, which further improves the results. 
For the processing pipeline, we integrated the NLP-based English detector operating on Czech and English ASR. The integrated English detector consists of TF-IDF for re-weighting the accumulated “soft” word counts and a logistic regression classifier to get the English/non-English decision~\cite{szoke21_interspeech}.

We created the development and evaluation dataset consisting of data from various airports, data with various English accents, and code-mixing of English and local languages. The data is selected from our ATCO2 corpus introduced in Table~\ref{tab:databases}. The development set is used to estimate the model parameters of our English language detector (the logistic regression classifier). The evaluation set is used for testing. The rules for manually labeling the utterances are mentioned below. We found several interesting properties of the ATC data during listening and tagging the ELD dataset:

\begin{itemize}
  \item Various noise conditions. The majority of data is clean, but there are some very noisy segments.
  \item Strongly accented English. The speakers’ English accent varies widely. From native speakers (pilots) to international accents (French, German, Russian, etc.) (pilots and ATCos) and strong Czech accents (pilots and ATCos).
  \item Mixed words and phrases. For example, the vocabulary  of Czech ATCos is a mix of Czech and English words. They use standard greetings in Czech which can be a significant portion of an "English" sentence if a command is short. On the other hand, they use many English words (alphabet, some commands) in "Czech" sentences. Moreover, they use a significant set of “Czenglish” words.
\end{itemize}

We use the language of spoken numerals as a rule of thumb to decide on the language of a particular ATCo-pilot communication utterance. The language has to be consistent within the audio recording. More detailed information including experimental results is covered in our previous work~\cite{szoke21_interspeech}.

\section{Conclusions}
\label{sec:conclusion}

This paper expands upon our previous work~\cite{zuluaga2022atco2} and discusses the main lessons learned from the ATCO2 project. The aim of the ATCO2 project was to develop a platform for collecting, pre-processing, and pseudo-annotating ATC communications audio data. With over 5000 hours of pseudo-annotated audio data, ATCO2 is the largest public ATC dataset to date. This project has pushed the research on robust automatic speech recognition and natural language understanding of ATC communications.

The main lessons learned from ATCO2 are as follows:

\begin{itemize}
    \item \textbf{Lesson 1:} ATCO2's automatic transcript engine (see Appendix~\ref{appendix:transcription-pipeline}) and annotation platform (see Appendix~\ref{appendix:anno-data-flow}) have proven to be reliable for collection of a large-scale audio dataset targeted to ATC communications. 
    
    \item \textbf{Lesson 2:} good annotation practices for ATC communications have been developed based on ontologies published by previous projects~\cite{helmke2018ontology}. A cheat-sheet (see Appendix~\ref{appendix:cheat-sheet}) has been created to provide guidance for future ATC projects and reduce confusion while generating transcripts.

    \item \textbf{Lesson 3:} training ASR systems purely on ATCO2 datasets (e.g., \textit{ATCO2-PL-500h set corpus}) can achieve competitive WERs on ATCO2 test sets (see Table~\ref{tab:asr-results}). The ASR model can achieve up to 17.9\%/24.9\% WERs on ATCO2-test-set-1h/ATCO2-test-set-4h, respectively. More importantly, these test sets contain noisy accented speech, which is highly challenging in standard ASR systems.
    
    \item \textbf{Lesson 4:} ATC surveillance data is an optimal source of real-time information to improve ASR outputs. The integration of air surveillance data can lead to up to 11.8\% absolute callsign WERs reduction, which represents an amelioration of 62.6\% (no boosting) $\rightarrow$ 82.9\% (GT boosted) of callsign accuracy in ATCO2-test-set-4h, as shown in Table~\ref{tab:atco2-boosting}.
    
    \item \textbf{Lesson 5:} ATCO2 corpora can be used for natural language understanding of ATC communications. BERT-based NER and speaker role detection modules have been developed based on ATCO2-test-set-4h. These systems can detect callsigns, commands, and values from the textual inputs. Additionally, speaker roles can also be detected based on textual inputs. The NLU task is of special interest to the ATC community because this high-level information can be used to assist ATCos in their daily tasks, thus, reducing their overall workload.
\end{itemize}

In conclusion, the lessons learned from the ATCO2 project and its methodology for collecting and pre-transcribing large-scale audio databases can be applied to other applications where data scarcity is a latent challenge.

\section*{Acknowledgements}

This paper presents a continuation of our previous work~\cite{zuluaga2022atco2}, the \textit{ATCO2 Corpus}. ATCO2 is derived from a joint contribution from Clean Sky 2 Joint Undertaking (JU) and EU-H2020, which was supported by Clean Sky 2 Joint Undertaking (JU) and EU-H2020, under Grant Agreement No. 864702—ATCO2 (Automatic collection and processing of voice data from air-traffic communications). See our website in \url{https://www.atco2.org/}.

\section*{Author Contributions}

Conceptualization, Juan Pablo Zuluaga Gomez, Srikanth Madikeri, Igor Szoke, Vincent Lenders, Mickael Rigault and Khalid Choukri; Data curation, Juan Pablo Zuluaga Gomez and Iuliia Nigmatulina; Formal analysis, Juan Pablo Zuluaga Gomez and Igor Szoke; Funding acquisition, Petr Motlicek; Investigation, Juan Pablo Zuluaga Gomez; Methodology, Juan Pablo Zuluaga Gomez, Iuliia Nigmatulina, Amrutha Prasad and Allan Tart; Project administration, Petr Motlicek; Resources, Juan Pablo Zuluaga Gomez; Software, Juan Pablo Zuluaga Gomez, Srikanth Madikeri, Allan Tart and Igor Szoke; Supervision, Petr Motlicek, Srikanth Madikeri, Igor Szoke and Vincent Lenders; Validation, Juan Pablo Zuluaga Gomez, Amrutha Prasad and Igor Szoke; Visualization, Juan Pablo Zuluaga Gomez and Mickael Rigault; Writing – original draft, Juan Pablo Zuluaga Gomez and Iuliia Nigmatulina; Writing – review \& editing, Juan Pablo Zuluaga Gomez, Iuliia Nigmatulina, Amrutha Prasad, Petr Motlicek, Driss Khalil, Vincent Lenders, Mickael Rigault and Khalid Choukri.

\section{Bibliography}
\bibliography{biblio}

\pagebreak
\appendix

\section{Automatic Transcription Engine}
\label{appendix:transcription-pipeline}

This appendix describes in details how we collected the audio and metadata that brought to live the \textit{ATCO2 corpus}. We mainly rely on the automatic transcription engine, described in more details in Section~\ref{subsec:data-collection-pipeline}. The automatic transcription engine is implemented as a scalable cloud service. It communicates with other services (or partners) using APIs. This service is designed to process large flows of data produced by data feeders.\footnote{Enthusiasts that act as 'feeders' of ATC speech and contextual ATC data (surveillance). See Section~\ref{subsec:transcriber}.}

\begin{wrapfigure}{R}{0.6\linewidth}
    \begin{center}
        \includegraphics[width=0.95\linewidth]{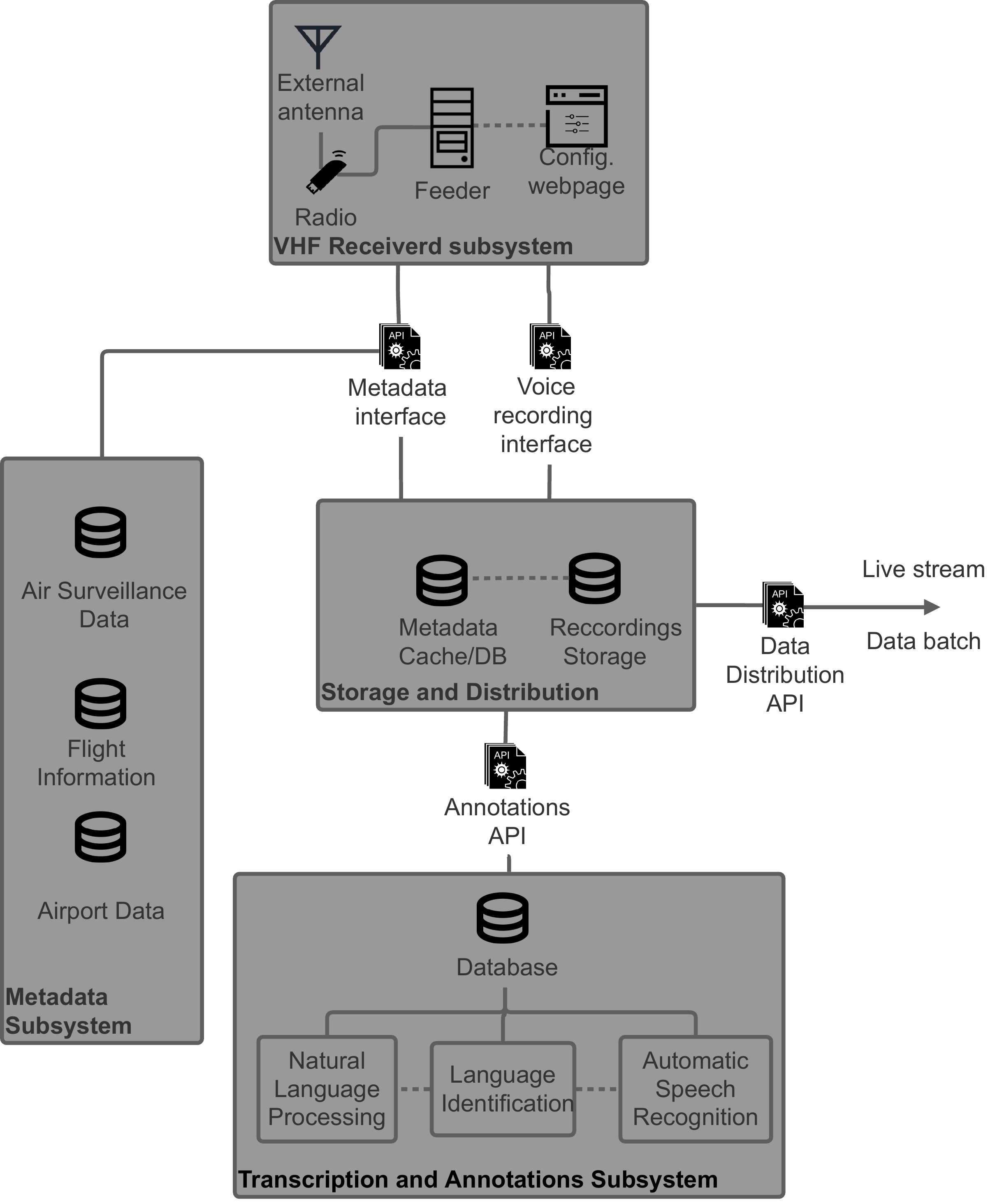}
        \caption{\textit{Overall ATCO2 communication schema}.}
        \label{fig:full_collection_pipeline}
    \end{center}
\end{wrapfigure} 

The data is pushed to this service by OSN\footnote{OpenSky Network: \url{https://opensky-network.org/}.} servers by calling an API request and providing a job setting JSON file. After the request is accepted, settings parameters are processed and the job is stored in an internal queue for processing. The user (in this case, OSN) may have an ability to tweak the settings and to affect the processing pipeline and the result. Namely:

\begin{itemize}
    \item Audio input format choices;
    \item Rejection threshold for too long audios;
    \item Rejection threshold for too short audios;
    \item Rejection threshold for too noisy audios;
    \item Rejection threshold for non-English audios;
    \item Switching the language of automatic speech recognizer.
\end{itemize}

Most of these are actually disabled due to security reasons (not to interrupt the processing pipeline), but may be easily enabled on the fly if needed. The overall data flow model is described in Figure~\ref{fig:full_collection_pipeline}. 
Any new job (request for a full automatic annotation of recording) accepted via API on the SpokenData\footnote{Industrial partner: \url{https://www.spokendata.com/atco2}.} side is processed by a master processing node. The job is enqueued into a workload manager queue. Once there is a free processing slot, the job is submitted to a processing server, or worker. The master processing node then informs the OSN server about the state of the job by calling a callback.

\pagebreak
\section{Annotation Platform: Data Flow}
\label{appendix:anno-data-flow}

The data (the recording for human annotation) lifecycle is split into 4 main states:

The new recording state is set as Queued and untouched when the recording is pushed into the annotation platform from the Transcription Engine. The recording is put into a queue of annotation jobs and is immediately visible to all annotators. The queue is shown in the Open jobs screen. Annotators can interact with the queue - listen to recordings and select some for the annotation. Recording in this state may drop off the queue in the case: they are old – no one is interested in annotating them, 3 annotators marked the recording by thumb down. The dropped off recordings are deleted after 7 days.

\begin{figure}[h!]
    \centering
    \includegraphics[width=0.9\linewidth]{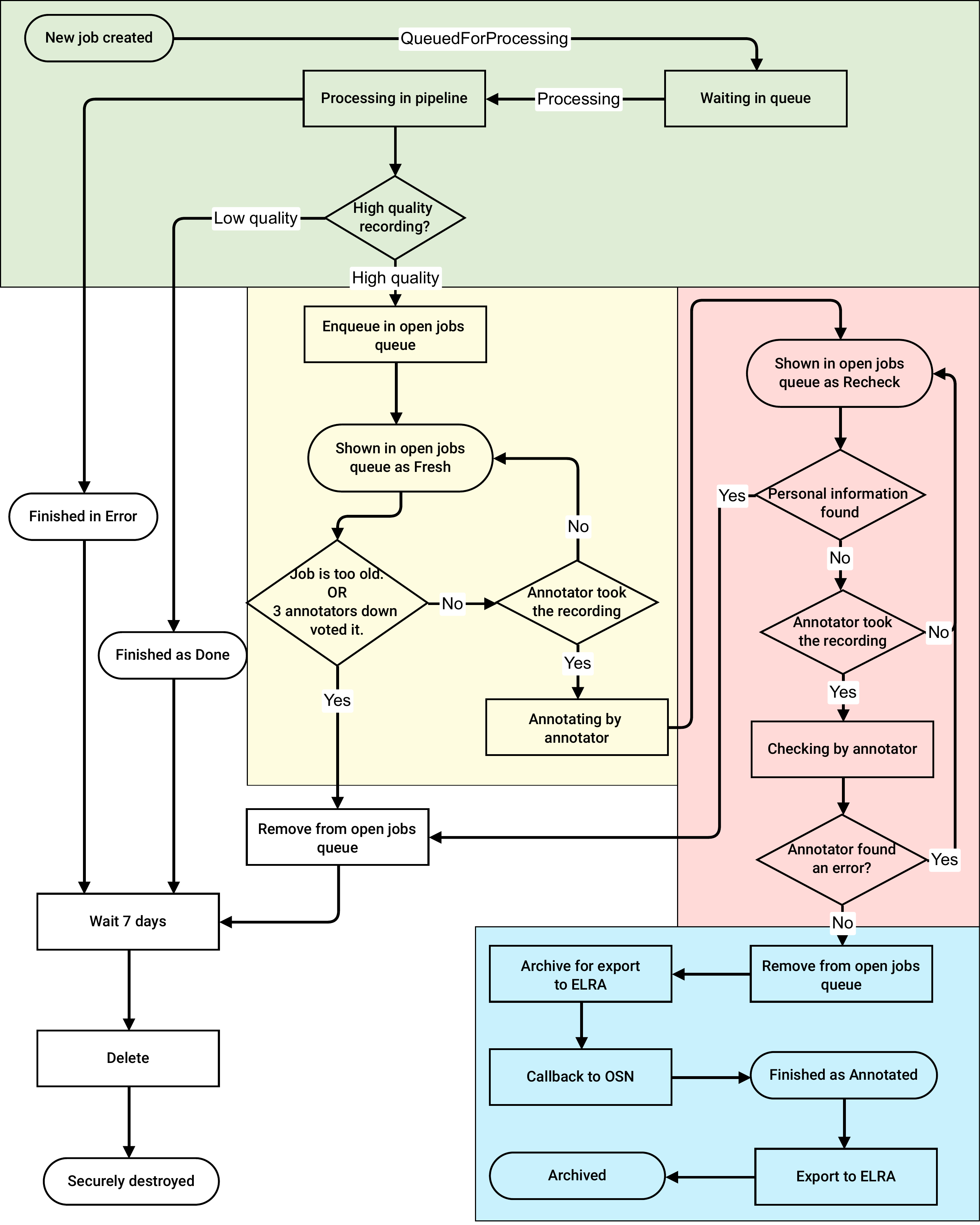}
    \caption{\textit{Data Flow of the annotation platform. Diagram of the data flow (lifetime) in the annotation platform.} Transcription engine in green. Queued and untouched state in yellow. Queued and annotated state in red. Annotated state in blue. The rest (white) is for state, Securely destroyed.}
    \label{fig:data_flow_annotation}
\end{figure}

Once a user selects the recording for annotation, revises the automatic annotation and saves it, the recording is set as Queued and annotated. This state prevents the recording from being dropped from the queue and deleted. Also, it is indicated as (to) Re-check in the Open jobs screen to inform other annotators, that it was modified (annotated) and they should re-check if the annotation is correct rather than annotate from scratch. If any annotator indicates the existence of personal information in the recording (by “Anonymize” label), the recording is dropped off the queue and deleted.

Next state is annotated. If the recording is successfully re-checked, then the recording is considered as annotated and the annotation is final. The recording is removed from the Open jobs queue and put on a stack of finished recording annotations. The stack is periodically exported to ELRA for further packaging and distribution to the community. This state also triggers a callback to OSN platform informing them that the human annotation is done, and they can download the transcription. After the recording was exported to ELRA, we set the state as \textit{\textbf{Finished}}. Here, the recording can be archived or deleted.
The detailed data flow schema is depicted in Figure~\ref{fig:data_flow_annotation}.

\pagebreak
\section{Communication Schema}
\label{appendix:communication-schema}

\begin{figure}[h!]
    \centering
    \includegraphics[width=0.9\linewidth]{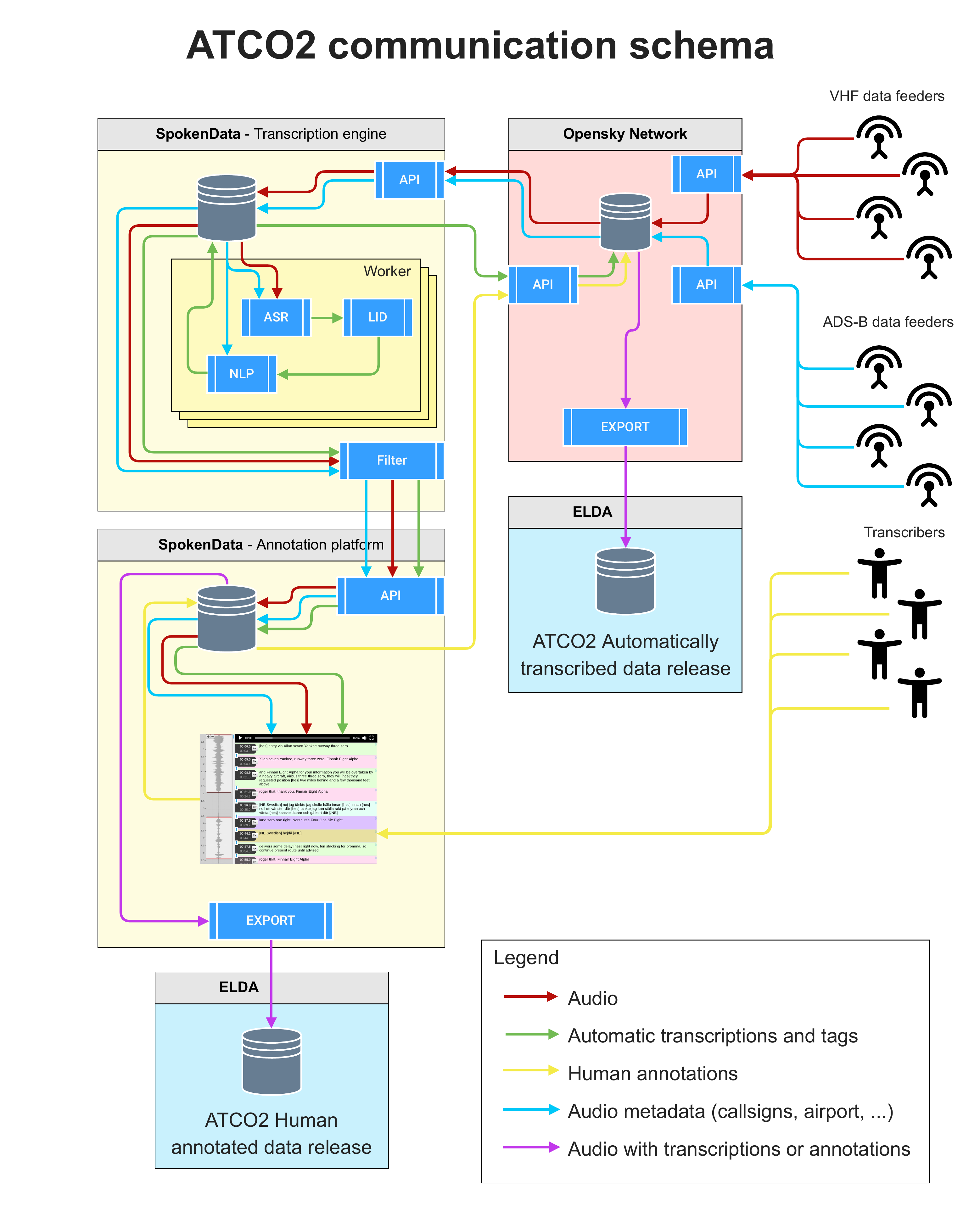}
    \caption{\textit{Overall ATCO2 communication schema.}}
    \label{fig:communication-schema}
\end{figure}

The communication schemar developed during ATCO2 project is depicted in Figure~\ref{fig:communication-schema}.

\pagebreak
\section{Annotation Cheat Sheet}
\label{appendix:cheat-sheet}

\begin{figure}[h!]
    \centering
    \includegraphics[width=0.9\linewidth]{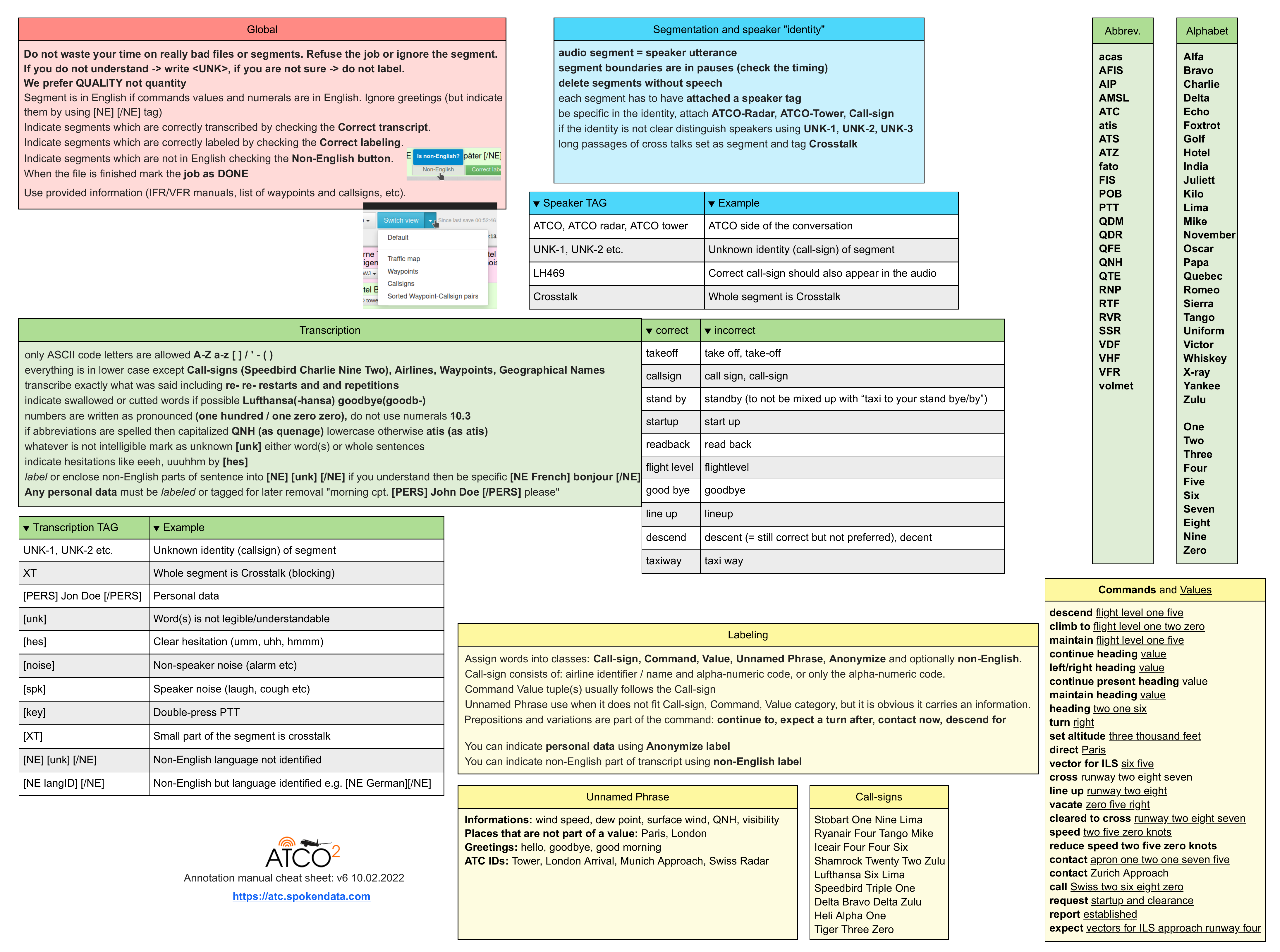}
    \caption{\textit{Cheat Sheet.} Document created during the annotation process of ATCO2 corpora. This document can be used to annotate air traffic control communications data from different airports.}
    \label{fig:cheat-sheet}
\end{figure}

Figure~\ref{fig:cheat-sheet} presents the annotation cheat sheet developed by ATCO2 project. 

\end{document}